\documentclass[twocolumn,english,nofootinbib,bibnotes,prb]{revtex4-1}

\usepackage[version=3]{mhchem}

\usepackage{graphicx}
\usepackage[usenames,dvipsnames]{color}
\usepackage{dcolumn}
\usepackage{bbm}
\usepackage{bm}
\usepackage{amsmath}
\usepackage{amssymb}
\usepackage{times}
\usepackage{placeins}
\usepackage{hyperref}
\hypersetup{colorlinks=false}
\usepackage{csquotes}
\usepackage{float}
\usepackage{multirow}

\newcommand{\nodagger}{{\phantom\dagger}}

\newcommand{\p}[1]{\ensuremath{\mathbf{#1}}}

\newcommand{\m}[1]{\ensuremath{\mathrm{#1}}}

\begin{document}
%------------------------------------

%------------------------------------
%\title{Cluster functional renormalization group and application to the $J_1$-$J_2$-Heisenberg model}
\title{Cluster functional renormalization group\\ and absence of a bilinear spin liquid in the $J_1$-$J_2$-Heisenberg model}
%------------------------------------

\author{Dietrich Roscher}
\author{Nico Gneist}
\author{Michael M. Scherer}
\author{Simon Trebst}
\author{Sebastian Diehl}
\affiliation{Institute for Theoretical Physics, University of Cologne, 50937 Cologne, Germany}

\date{\today}

%------------------------------------
\begin{abstract}
The pseudofermion functional renormalization group (pf-FRG) has been put forward as a semi-analytical scheme that, 
for a given microscopic spin model, allows to discriminate whether the low-temperature states exhibit magnetic ordering or a tendency towards the formation of quantum spin liquids. However, the precise nature of the putative spin liquid ground state has remained hard to infer from the original (single-site) pf-FRG scheme.
Here we introduce a cluster pf-FRG approach, which allows for a more stringent connection between a microscopic spin model and its low-temperature spin liquid ground states. In particular, it allows to calculate spatially structured fermion bilinear expectation values on spatial clusters, which are formed by splitting the original lattice into several sublattices,
thereby allowing for the positive identification of a family of bilinear spin liquid states.
As an application of this cluster pf-FRG approach, we consider the $J_1$-$J_2$ SU($N$)-Heisenberg model on a square lattice, which is a paradigmatic example for a frustrated quantum magnet exhibiting quantum spin liquid behavior for intermediate coupling strengths. In the well-established large-$N$ limit of this model, we show that our approach correctly captures the emergence of the $\pi$-flux spin liquid state at low temperatures. For small $N$, where the precise nature of the ground state remains controversial, we focus on the widely studied case of $N=2$, for which we determine the low-temperature phase diagram near the strongly-frustrated regime after implementing the fermion number constraint by the flowing Popov-Fedotov method.
Our results suggest that the $J_1$-$J_2$-Heisenberg model does not support the formation of a fermion bilinear spin liquid state.
\end{abstract}

%------------------------------------

\maketitle

%------------------------------------
\section{Introduction}
\label{Intro}
%------------------------------------

Quantum magnets are host to an astounding range of fascinating phenomena which go beyond the realm of classical magnetism. This includes the formation of multipolar order, such as spin nematic states, or more generally topological order, along with the appearance of fractionalized excitations and long-range entanglement \cite{lacroix2011introduction}. The latter are hallmarks of quantum spin liquids (QSLs) \cite{balents2010spin}, which have been widely studied in the context of frustrated magnets. Substantial progress in the understanding of the emergence of such QSLs has been achieved through the seminal work by Kitaev~\cite{Kitaev2006}, providing an exactly solvable microscopic model with various QSL ground states. Aside from this particular model system, it remains, however, notoriously difficult to establish reliable connections between microscopic spin Hamiltonians and their possible QSL ground states. 

A paradigmatic example for the challenges to be met is the $J_1$-$J_2$ Heisenberg model. Here, spins on a two-dimensional square lattice interact antiferromagnetically with their nearest and next-to-nearest neighbors with couplings $J_1$ and $J_2$, respectively, and the level of frustration can be tuned by the ratio of the couplings $J_2/J_1$.
The apparent simplicity of this model combined with an early finding~\cite{Chandra1988} of a low-temperature non-magnetic phase in the strongly frustrated intermediate coupling regime has drawn the attention of researchers for over three decades now. 
Many methods have been developed and applied to clarify the precise nature of the frustrated ground state, including exact diagonalization~\cite{ED95,ED96,ED00,ED06,ED09}, density matrix renormalization group~\cite{DMRG12,DMRG13,DMRG17}, tensor network techniques~\cite{TPS16,PEPS17,PEPS18,LatestPEPS18}, variational approaches~\cite{Variational12,Variational13}, different expansion schemes~\cite{Chandra1988,SpinWave06,PlaqExp16} and others~\cite{MasterEq09,ClusterMF13,RG14,MC14,CC14}. 
For the ground state, gapped~\cite{DMRG12} and gapless~\cite{Variational13,CC14,MC14,TPS16,DMRG17} spin liquids have been found as well as different types of valence bond solids~\cite{ED00,ED06,SpinWave06,MasterEq09,DMRG13,RG14,PlaqExp16,PEPS18}, but the results are widespread and  not congruent within or across the applied approaches, leaving a rather unsatisfactory situation. 

The large range of incompatible results on the ground state of  the frustrated $J_1$-$J_2$ Heisenberg model suggests that it is rather sensitive to any bias introduced by an approximate approach. The goal of this work is to avoid  any such bias by construction.
We employ and expand renormalization group techniques, which have successfully been used in diverse physical contexts to systematically study effective low-energy models of microscopic theories. A particular realization, the pseudofermion functional renormalization group (pf-FRG) has the capacity to deal with strongly correlated spin systems and many works have proven its ability to accurately characterize quantum magnets in complicated lattice geometries with respect to ordering patterns as well as critical temperatures and couplings~\cite{Reuther2010,Reuther2011,Reuther2011c,Reuther2014,Buessen2016,Iqbal2016,BaezLargeS17,Buessen2017}.

Until recently, a central drawback of the pf-FRG approach has been that the occurrence of magnetized and even certain types of QSL ground states leads to an instability in the renormalization group flow. While this instability itself represents a hallmark of ordering, it prevents a continuation of the flow towards the low-energy regime. Therefore, the precise nature of the ground state or even competing orders remained elusive.
In Refs.~\onlinecite{LargeNReal,LargeNMomentum} it was demonstrated, that the pf-FRG can be properly regularized such that the occurrence of an instability does not lead to a breakdown of the description itself anymore. Since this result can be achieved with arbitrarily small (initial) bias, it can be used to positively identify and characterize spin-liquid phases that can be represented by expectation values of pseudofermion bilinears~\cite{Wen2002}, which we refer to as {\em bilinearly ordered} spin liquids in the following. %These 

In this work, we substantially extend this regularized pf-FRG approach by working with multi-site clusters in lieu of the single-site perspective employed hitherto. Importantly, this allows to map part of the spatial structure of spin-liquid order parameters onto the pseudofermion representation itself. This, in turn, enables us to systematically include more spin-liquid ordering patterns in the characterization of the low-energy physics. 
Employing this cluster pf-FRG approach to the $J_1$-$J_2$ SU($N$)-Heisenberg model on the square lattice, we show that benchmark calculations for the large-$N$ case indeed exhibit the expected emergence of a $\pi$-flux spin liquid state in the low-energy regime. This reveals the true power of the cluster pf-FRG approach:  the RG flow automatically and reliably chooses the energetically most favorable ordering pattern available at the given level of approximation. This distinct feature of the approach can not only be used to systematically improve the quality of predictions for spin-liquid phases, but statements about their very existence are found to be independent of the level of approximation employed for the spatial structure. Taking into account the full set of non-magnetic bilinear ordering patterns, the cluster pf-FRG approach thus allows us to obtain definite statements on the existence of magnetically ordered or spin-liquid phases in the finite-temperature phase diagram of the $J_1$-$J_2$ model.

%------------------------------------
\subsection{Line of Arguments and Overview of Results}
\label{Keyres}
%------------------------------------

We develop a generalization of the pf-FRG method to the $J_1$-$J_2$ Heisenberg model at zero and finite temperatures, aiming at a clarification of whether the frustrated regime can be described by a bilinear spin liquid.
First, in order to establish a well-controlled limit without having to deal with complicated fermion number constraints, we revisit the SU($N$) symmetric model at large $N$. We show that our generalization correctly captures the emergence of the $\pi$-flux spin liquid state at low energies, cf. Sec.~\ref{LargeN}, by introducing a splitting of the square lattice into four sublattices, cf. Sec.~\ref{Sub1_OP}, and an appropriate set of order parameters. In particular, we show that the RG flow automatically selects the energetically favored ground state.
This analysis corroborates the suitability of our approach to detect any (non-magnetic) bilinearly ordered spin liquid states despite the employed approximations.
Moreover, we clearly exhibit that the spin-liquid behavior at large $N$ results from the renormalization group flow of the four-fermion function and is not related to a frequency-dependent self-energy, Sec.~\ref{FreqDissLargeN}.
Introducing an artificial damping into the self-energy causes an unphysical shift of the critical temperature and order parameter.
With this knowledge, we turn to the physically relevant case of $N=2$, where we first carefully implement the fermion number constraint in terms of a flowing Popov-Fedotov method, Sec.~\ref{PopotovRulez}, and then calculate the low-temperature phase diagram as a function of the ratio $J_2/J_1$, Sec.~\ref{Ngleich2}.
We find that our approach predicts the appearance of magnetic instabilities through the full range of $J_2/J_1$ with 
only a moderate suppression in the strongly-frustrated regime.
In view of the insight on the role of frequency dependencies, 
this suggests that the $J_1$-$J_2$ Heisenberg model does not support the formation of a bilinearly ordered spin liquid state.
Conclusions and future prospects are given in Sec.~\ref{Cloudlook}.

%------------------------------------
\section{Model and Implementation}
\label{Basics}
%------------------------------------

The goal of our analysis is to characterize the ground state of the Heisenberg-Hamiltonian
\begin{equation}
\label{BaseHamil}
\mathcal{H} = \frac{J_1}{N}\sum_{\langle ij \rangle} \p{S}_i\cdot \p{S}_j + \frac{J_2}{N}\sum_{\langle\langle ij \rangle\rangle} \p{S}_i\cdot \p{S}_j,
\end{equation}
implemented on a two-dimensional square lattice. The nearest ($J_1$) and next-to-nearest ($J_2$) neighbor couplings are positive and thus favor antiferromagnetic correlations. Their ratio $g = J_2 /J_1$ is an important quantity, encoding the frustration of the system.

%------------------------------------
\subsection{Pseudofermion FRG}
\label{pf-FRG}
%------------------------------------

Our tool of choice is the pseudofermion formulation of functional renormalization group (FRG). At its heart lies the functional differential Wetterich equation~\cite{Wetterich1993}
\begin{equation}
\label{Wettereq}
\partial_\Lambda \Gamma_\Lambda = \frac{1}{2}\int_\tau\sum_i f^\dagger \partial_\Lambda \mathcal{P}_\Lambda f + \frac{1}{2}\m{STr}\left[\frac{\partial_\Lambda \mathcal{P}_\Lambda}{\Gamma_\Lambda^{(2)}}\right]\,,
\end{equation}
for the running effective action $\Gamma_\Lambda$ with multiplicative regularization~\cite{SalmHon2001}. Here, $\{f^\dagger,f\}$ are anticommuting Grassmann numbers.
At the initial scale $\Lambda\rightarrow\Lambda_{\m{UV}}$, $\Gamma_\Lambda$ is given by the microscopic action corresponding to the Hamiltonian~\eqref{BaseHamil}. After choosing a regulator function by which to multiply the free propagator $\mathcal{P}_\Lambda$, solving the flow equation, Eq.~\eqref{Wettereq}, in principle yields the full effective action $\Gamma_{\Lambda\rightarrow 0}$ from which thermodynamic properties can be read off.

Practically, two prerequisites are needed to actually perform this procedure. Firstly, the Hamiltonian operator~\eqref{BaseHamil} needs to be recast in a form that is amenable to our chosen formulation of the FRG flow Eq.~\eqref{Wettereq}. To that end, we choose the pseudofermion representation of spin operators~\cite{Abrikosov1965}
\begin{equation}
\label{PFDef}
S_i^\mu = f^\dagger_{i\alpha}T^\mu_{\alpha\beta} f^\nodagger_{i\beta}\,,
\end{equation}
where $T^\mu$ are matrices of the underlying spin symmetry group. Thus, $T^\mu = \frac{1}{2}\sigma^\mu$ for the physical SU(2) symmetry with $\sigma^\mu$ being Pauli matrices, and generalized Gell-Mann matrices for general SU($N$). To eliminate unphysical degrees of freedom, the (pseudo)fermions are also subject to the local constraint on the particle number,
\begin{equation}
\label{OpConstraint}
f^\dagger_{i\alpha}f^\nodagger_{i\alpha} = \frac{N}{2}\,\quad \forall i.
\end{equation}
Making use of commutation relations between the $T^\mu$ and applying the constraint to eliminate particle number operators, the Hamiltonian~\eqref{BaseHamil} can be recast into an action that serves as initial condition for the flow equation,
\begin{equation}
\label{BareS}
\begin{aligned}
S &= \int_\tau\left[\sum_i f^\dagger_{i\alpha}\partial_\tau f^\nodagger_{i\alpha} \right.\\
&\left.- \frac{J_1}{N}\sum_{\langle ij \rangle} f^\dagger_{i\alpha}f^\nodagger_{j\alpha}f^\dagger_{j\beta}f^\nodagger_{i\beta} - \frac{J_2}{N}\sum_{\langle\langle ij \rangle\rangle} f^\dagger_{i\alpha}f^\nodagger_{j\alpha}f^\dagger_{j\beta}f^\nodagger_{i\beta}\right]\,,
\end{aligned}
\end{equation}
where, in the following, the explicit spin indices will often be suppressed for better readability. %conciseness. 

We note that  the flow equation \eqref{Wettereq} can usually not be solved without an approximate ansatz for $\Gamma_\Lambda$. The choice of this ansatz is crucial since all structures relevant to the physics of interest need to be included, while keeping the computational cost manageable. Systematic development and benchmarking of this truncation for $\Gamma_\Lambda$ is a central objective of our analysis in the forthcoming sections.

%------------------------------------
\subsection{Sublattice representation: Spatially structured orders}
\label{Sub1_OP}
%------------------------------------

The inclusion of explicit flowing order parameters in the pf-FRG scheme by construction reduces the symmetry of the problem and thus increases computational complexity. For this reason, a point-like momentum-space projection 
was applied in Ref.~\onlinecite{LargeNMomentum} instead of the conventional spatially resolved pf-FRG schemes~\cite{Reuther2010}. In this projection scheme, $J_{1,2}$ does not depend on frequency or momentum. This approach trades spatio-temporal resolution for explicit information on ordering tendencies and access to the physically relevant low-energy regime. Consequently, only a spatially homogeneous Baskaran-Zou-Anderson (BZA)-phase could be detected in the large-$N$ analysis so far, while it is known~\cite{AffleckMarston88,ArovasAuerbach88} that (spatially structured) $\pi$-flux or even Peierls phases are energetically preferred.

For the analysis of possible spin liquid phases in the $J_1$-$J_2$ model, this seems too restrictive, as it would exclude the majority of possible spin-liquid ordering patterns~\cite{Wen2002}. 
Furthermore, the next-to-nearest neighbor interaction $\sim J_2$ needs to be included and discriminated from the $\sim J_1$ structure.
In order to discriminate the $J_1$- and $J_2$-coupling structures or different spatially-structured ordering,  
one could abandon the simple point-like projection in momentum space in favor of an actual, finite discretization of the Brillouin zone. This would increase computational cost to a degree approaching or even exceeding that of the conventional pf-FRG scheme.

\begin{figure}[t!]
\includegraphics[width=0.75\columnwidth]{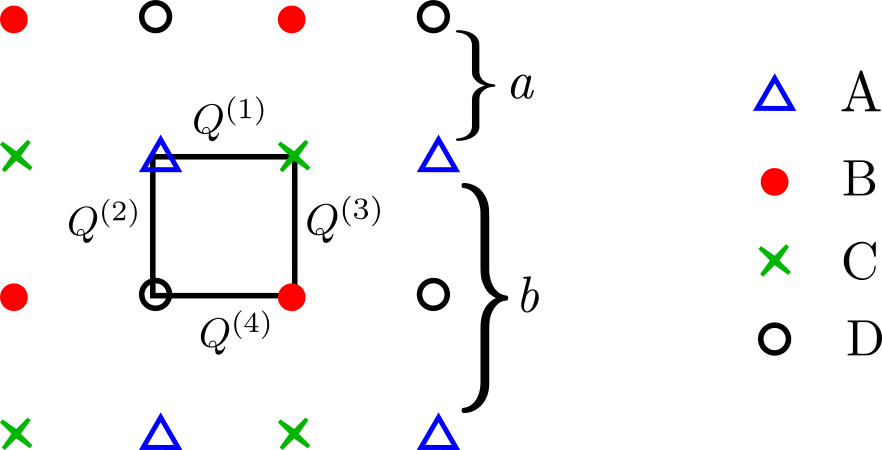}
\caption{Four-sublattice representation of the 2D square lattice with inequivalent order parameters (link variables) $Q^{(1..4)}$. $a$ and $b$ are the lattice constants of the original and the enlarged unit cells, respectively.}
\label{FourLattice}
\end{figure}

We develop a different approach here, where the point-like projection scheme is kept and discrimination between structures on different bonds can be performed algebraically. When representing the nearest-neighbor structure of the $J_1$ interaction, it is natural to artificially split the original square lattice into two equivalent sublattices $A$ and $B$, defining the interaction on the links between them~\cite{ArovasAuerbach88}. We perpetuate this idea by moving to \emph{four} artificial sublattices, see Fig.~\ref{FourLattice}	. Spinors can now be defined on the corresponding four-dimensional space:
\begin{equation}
\label{SpinorDef}
\Psi^\dagger = \left(f^{\dagger,A},f^{\dagger,B}\right) \,\rightarrow\, \Psi^\dagger = \left(f^{\dagger,A},f^{\dagger,B},f^{\dagger,C},f^{\dagger,D}\right).
\end{equation}
By properly choosing $4\times 4$ matrices $\eta^{X/Y}$, the general interaction structure in frequency-momentum space,
\begin{equation}
\label{SubWWStruc}
\begin{aligned}
&\frac{J_n}{N}\int_{p_1..p_4\in \m{TZ}}\left(\Psi^\dagger_{p_1}\eta^X_n\Psi^\nodagger_{p_2}\right)\left(\Psi^\dagger_{p_3}\eta^Y_n\Psi^\nodagger_{p_4}\right)\\
&\qquad\cdot f(p_1,p_2,p_3,p_4)\cdot\delta_{p_1p_2p_3p_4}\,,
\end{aligned}
\end{equation}
can now be restricted to nearest- or next-to-nearest neighbors. Note that we have suppressed Matsubara labels. Further, TZ defines the \enquote{tiny zone}, corresponding to one quarter of the Brillouin zone of the original lattice. The geometry-induced momentum structure is represented by $f(p_1,p_2,p_3,p_4)$ and $\delta_{p_1p_2p_3p_4}$ is a momentum-conserving $\delta$-function.

%------------------------------------
%------------------------------------

Using the representation in Eq.~\eqref{SubWWStruc} it is possible to resolve spatial patterns of intermediate range in a simple algebraic and systematic way without having to consider the explicit momentum dependence of interaction vertices and self-energies. One can, in principle, always increase the number of artificial sublattices to increase spatial range. Furthermore, this strategy can be generalized to other lattice geometries. For example, a 2D Kagome lattice may be described as a triangular lattice, split into four sublattices, one of which is eliminated~\cite{KagomeConstruct95}. This will be the subject of future work.
We also note that the present scheme allows for a comparatively simple algebraic analysis of spatial symmetry breaking patterns. This aspect will be discussed in detail in Sec.~\ref{Sub2_FancyGroups}, below.

The interaction structure given by the Hamiltonian~\eqref{BaseHamil} with respect to the sublattice definition from Eq.~\eqref{SpinorDef} can now be described uniquely by the non-vanishing entries of the respective $\eta^{X/Y}$ matrices, see Tab.~\ref{InitWWStruc}.
\begin{table}[H]
\centering
\begin{tabular}{|c|c|c|c|}
\hline
 & & & \\[-1em]
Coupling & $\eta^X$ & $\eta^Y$ & $f(p_1,p_2,p_3,p_4)$\\
\hline
$J_{1,\Lambda}$ & 13 & 31 & $\cos(p_{2,x} - p_{3,x})$\\
$J_{1,\Lambda}$ & 14 & 41 & $\cos(p_{2,y} - p_{3,y})$\\
$J_{1,\Lambda}$ & 23 & 32 & $\cos(p_{2,y} - p_{3,y})$\\
$J_{1,\Lambda}$ & 24 & 42 & $\cos(p_{2,x} - p_{3,x})$\\
\hline
$J_{2,\Lambda}$ & 12 & 21 & $\cos(p_{2,x} - p_{3,x})\cos(p_{2,y} - p_{3,y})$\\
$J_{2,\Lambda}$ & 34 & 43 & $\cos(p_{2,x} - p_{3,x})\cos(p_{2,y} - p_{3,y})$\\
\hline
\end{tabular}
\caption{Momentum structure and non-vanishing entries of the sublattice-space matrices $\eta^{X/Y}$ for the initial interaction of the $J_1$-$J_2$ model.}
\label{InitWWStruc}
\end{table}
The geometric momentum structure follows directly from the spatial information contained in the $\eta^{X/Y}$ matrices. 
Here, we exclusively consider couplings between the nearest available sublattice sites for a given $\eta^{X/Y}$-matrix. This amounts to an approximation, which will be discussed in Sec.~\ref{Pairing}, since for finite $N$, longer-ranged contributions to the respective vertices are generated.
We disregard those to keep the system of RG equations closed.

%------------------------------------
\section{Large-$N$ analysis}
\label{LargeN}
%------------------------------------

As a first step towards a systematic study of the full model, Eq.~\eqref{BaseHamil}, we perform a large-$N$ analysis.
This provides useful hints on the character of a possible spin liquid phase in the frustrated regime and guides the more intricate search at the physical value of $N=2$. Furthermore, for large $N$, it is sufficient to implement the fermion number constraint~\eqref{OpConstraint} {\em on average} only~\cite{ArovasAuerbach88}, which greatly simplifies the analysis.

We consider the well-understood limit $J_2 = 0$ first: To regularize divergences in the RG flow of the running coupling $J_{1,\Lambda}$, we introduce bilinear density-type symmetry-breaking order parameters $Q^{(1..4)}$. These may now be addressed in a manner analogous to Eq.~\eqref{SubWWStruc},
\begin{equation}
\label{SubQStruc}
Q^{(n)}\int_{p\in\m{TZ}} \left(\Psi^\dagger_p\eta^X_n\Psi^\nodagger_p\right) f(p)\,.
\end{equation}
Possible order parameters compatible with four sublattices are given by their matrix structure in Tab.~\ref{InitQStruc}.

%--------------------------------------
\begin{table}[H]
\centering
\begin{tabular}{|c|c|c|}
\hline
 & & \\[-1em]
Order Parameter & $\eta^X$ & $f(p)$\\
\hline
$Q^{(1)}$ & \{13,31\} & $\cos(p_x)$\\
$Q^{(2)}$ & \{14,41\} & $\cos(p_y)$\\
$Q^{(3)}$ & \{23,32\} & $\cos(p_y)$\\
$Q^{(4)}$ & \{24,42\} & $\cos(p_x)$\\
\hline
$Q^{(5)}$ & \{12,21\} & $\cos(p_x)\cos(p_y)$\\
$Q^{(6)}$ & \{34,43\} & $\cos(p_x)\cos(p_y)$\\
\hline
\end{tabular}
\caption{Momentum structure and non-vanishing entries of the sublattice-space matrix $\eta^{X}$ for density-type order parameters.}
\label{InitQStruc}
\end{table}
%--------------------------------------

At large-$N$, pairing-type order parameters $\Delta$ are suppressed and will therefore not be considered here. When we proceed to $N=2$ below, we will take them into account, though.
Moreover, since we are primarily interested in identifying spin-liquid states, we will not introduce magnetic order parameters $\sim M^{(n)} \Psi^\dagger_\alpha \tau_{\alpha\beta}\eta^X\Psi_\beta$ where $\tau_{\alpha\beta}\neq \delta_{\alpha\beta}$ in spin space, neither for $N=2$ nor for large $N$. This encompasses all possible bilinear order parameters.
Magnetic phases will therefore be signaled by a divergence of the RG flow. In contrast, bilinearly-ordered spin-liquid states are identified by a regular RG flow with a finite order parameter in the infrared.

Due to the presence of the symmetry-breaking order parameters, new interaction structures besides the initial ones in Tab.~\ref{InitWWStruc} are generated during the flow. In the space of four sublattices, up to $4^2\cdot 4^2 = 256$ different structures can in principle occur and may be discriminated algebraically. However, not all of these are finite in the present setup. It can be shown that only matrices $\eta^{X/Y}$ which are already present in the initial $\Gamma_{\Lambda\rightarrow\Lambda_{\m{UV}}}$ are available for combination among themselves at $N\rightarrow\infty$, cf. App.~\ref{NFlucMat} for details. This reduces the number of newly generated interaction structures down to 36. Some of theses are related by hermitean conjugation, which can be used to further reduce the number of differential equations.
The resulting 40 coupled differential equations are rather complex. We therefore refrain from writing them down explicitly. Determining potentially non-vanishing couplings as well as the construction of the flow equations themselves follows a set of comparatively simple and straightforward rules. Both problems are therefore amenable to automated and/or numerical analysis. 

%------------------------------------
\subsection{Staggered-flux spin liquid ground state}
\label{piflux}
%------------------------------------

\begin{figure}[t!]
\includegraphics[width=\columnwidth]{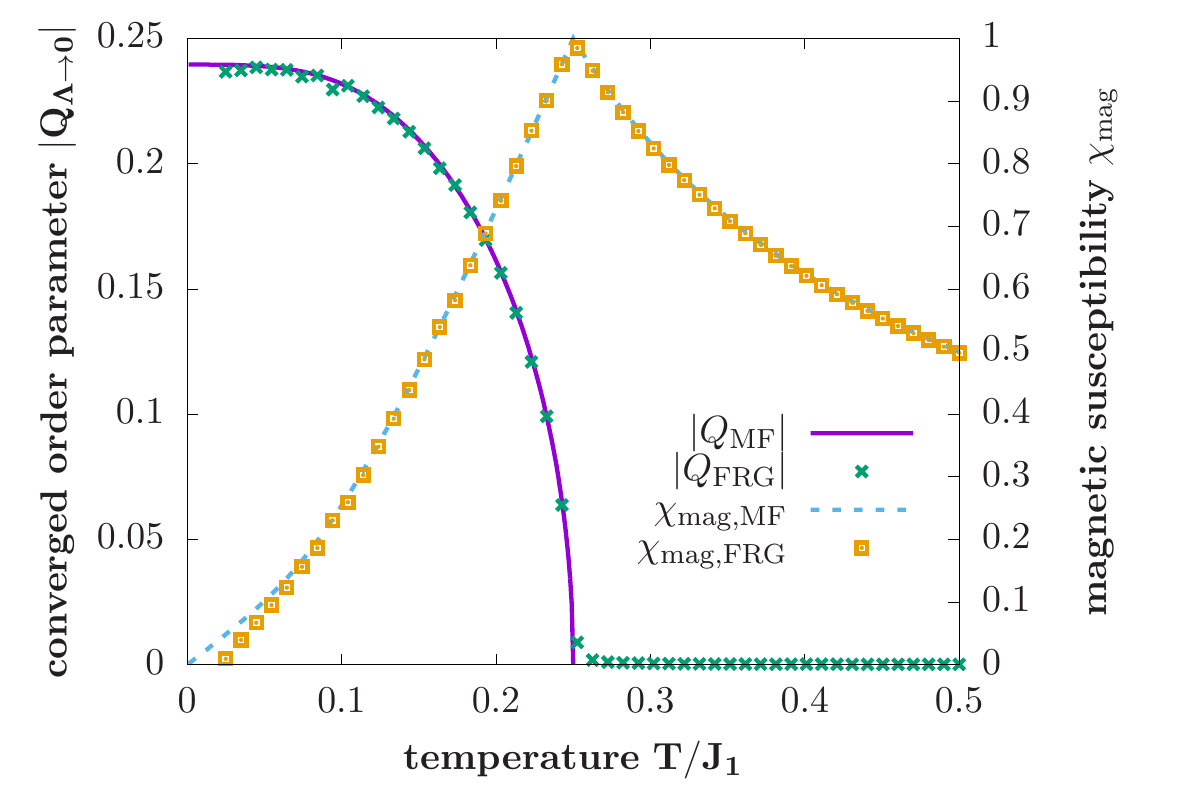}
\caption{Absolute values of the order parameter $Q$ and magnetic susceptibility $\chi_{\rm{mag}}$ from mean field (solid/dashed lines) and four-sublattice pf-FRG (crosses/boxes). The deviations at low temperature are due to the limited number of Matsubara frequencies taken into account.}
\label{PiFluxObsPlot}
\end{figure}

\begin{figure}[t!]
\includegraphics[width=\columnwidth]{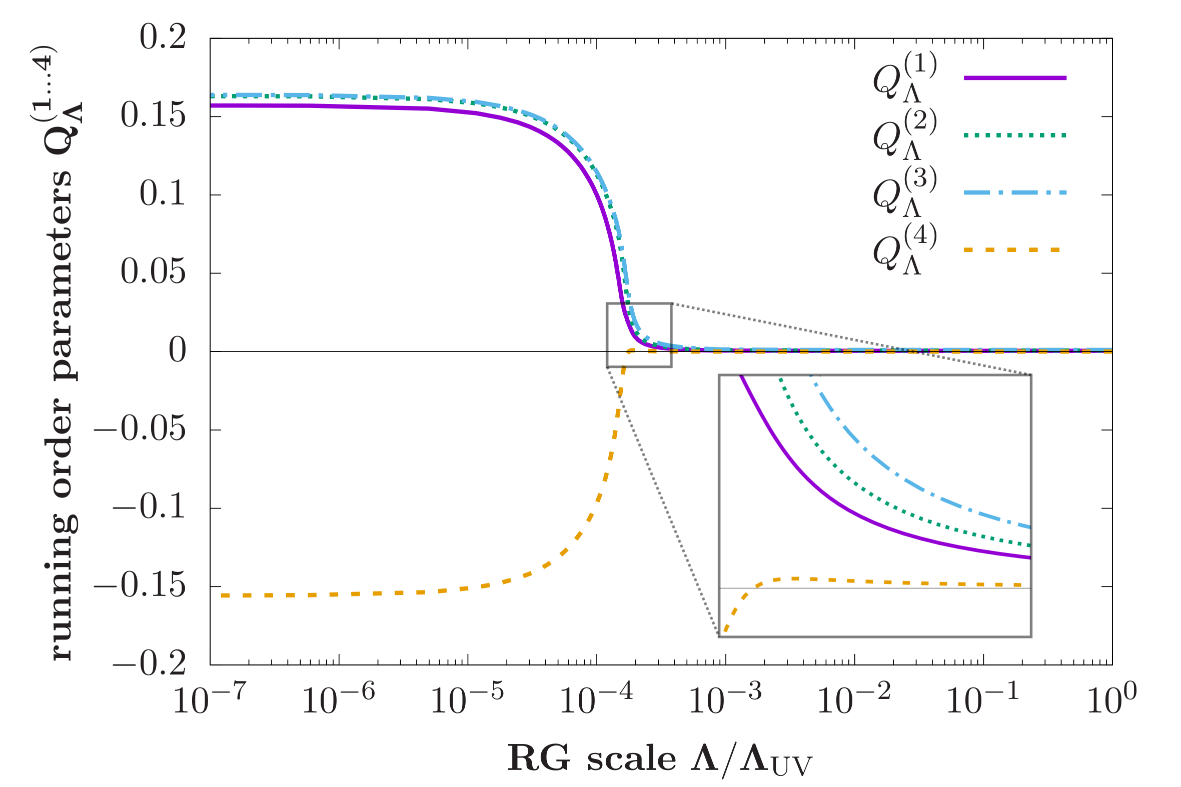}
\caption{Link-dependent values of the order parameter, cf. Tab.~\ref{InitQStruc}, of a random set of real initial values for the $Q^{(1..4)}_{\Lambda\rightarrow\Lambda_{\rm{UV}}}$. Up to numerical inaccuracies their infrared values are $Q^{(1)} = Q^{(2)} = Q^{(3)} = -Q^{(4)}$. Therefore, the complex phase between the four variables surrounding one plaquette is $\pm \pi$. The inset shows that the $\pi$-flux structure is enforced by the RG flow although the employed set of initial values for $Q_\Lambda^{(1..4)}$ is strictly positive.}
\label{FluxQPlot}
\end{figure}

For large $N$, the mean-field approach to the SU($N$) Heisenberg model becomes exact and a staggered- or $\pi$-flux spin liquid phase arises as the ground state~\cite{AffleckMarston88,Wen2002}.
The large-$N$ FRG flow equations are known to exactly reproduce the mean-field results if the Katanin scheme~\cite{KataninSeminar04} is employed, as shown previously~\cite{Salmhofer2004,LargeNMomentum}. Initializing the $Q^{(1..4)}$ with small $(\mathcal{O}(10^{-4}))$, generally complex, random numbers, we find excellent agreement for the absolute value of the order parameters as well as the magnetic susceptibility $\chi_{\m{mag}}$ with the mean-field solution,
see Fig.~\ref{PiFluxObsPlot}. The specific values of the individual order parameters themselves do depend on the initial values, see Fig.~\ref{FluxQPlot} for an example, in the sense that different but equivalent configurations arise: Except for one special configuration, see Sec.~\ref{Sub2_FancyGroups}, we always find an overall phase of $\pi$ around each plaquette as expected. The RG flow thus automatically selects the appropriate state.

We conclude that the sublattice splitting is indeed a suitable method to systematically capture spatial structures of spin liquid phases. A splitting into eight sublattices should give access to the true large-$N$ ground state which is known to be a Peierls phase~\cite{AffleckMarston88}. We refrain from pursuing this, here, and focus on the $N=2$ case, in the following.

%------------------------------------
\subsection{Algebraic ordering analysis}
\label{Sub2_FancyGroups}
%------------------------------------

Xiao-Gang Wen's symmetry-based classification of spin liquid states relies on a decoupling of fermion bilinears in terms of order-parameter-like mean-fields~\cite{Wen2002} which could, in principle, be related to the fermionic interaction by a Hubbard-Stratonovich transformation.
In our fermionic RG formulation of spin models, such ordering tendencies manifest themselves in a divergence of the fermionic couplings~\cite{Braun2012,LargeNReal} -- unless regularized.
This divergence is crucial for our approach, as it guarantees that this type of bilinear ordering cannot be missed. 
However, an order parameter which regularizes the fermionic RG flow, allowing for its continuation towards the infrared, may still not reflect the true symmetry-breaking pattern of the ground state.
For instance, including the homogeneous BZA state in a simple two-sublattice setup 
leads to a convergent RG flow and no sign of something being amiss could be discerned.
On the other hand, the ground state of the system is better described by the $\pi$-flux phase, as described in the previous section, which is characterized by an additional breaking of translation invariance.
The difference between the two states lies in the spatial structure of the bilinear expectation value $Q_{ij} \sim \langle f^\dagger_i f^\nodagger_j\rangle$, serving as an order parameter for translational/rotational symmetry breaking. The occurrence of this spatial structure by itself is not a priori related to a diverging coupling and can thus be missed by a na\"ive implementation.
Considering the importance of additional breaking of discrete as well as non-compact symmetries %like translation invariance 
for the classification of spin liquid states~\cite{Wen2002}, this is an issue that will therefore be addressed in the following.%appears to be a huge drawback of the method.

Employing the cluster pf-FRG introduced above, this situation is remedied in a systematic way. The square-lattice Heisenberg model is symmetric under primitive translations with respect to the lattice spacing $a$ along $\vec{e}_x$ and $\vec{e}_y$ and a number of other symmetries, forming a discrete subgroup of the non-compact Galilei group associated with non-relativistic continuum systems. When introducing a number $M$ of sublattices, part of this group is mapped onto a subgroup of the compact SU($M$)\footnote{As long as no pairing order parameters $\Delta\sim f_{i\alpha} f_{i\beta }\epsilon_{\alpha\beta}$ are introduced, the action may even be invariant under U($M$). The generator $\mathbbm{1}_{4\times 4}$ is, however, not associated with spatial symmetry breaking. We therefore do not take it into account and consider SU($M$) as our starting group instead.} that transforms the spinor $\Psi$ from Eq.~\eqref{SpinorDef} where, for instance, $M=4$. Moving from sublattice $A$ to $B$ would be described by a matrix $\eta^{12}\in$ SU$(4)$ instead of an element of the Galilei group. The latter is reduced to describe translations with twice the original lattice spacing, $b=2a$. Consequently, translational symmetry breaking on the scale of single plaquettes is now characterized by spontaneous breaking of the subgroup of SU$(4)$ that is compatible with the initial interaction given in Tab.~\ref{InitWWStruc}.

Let us make this more explicit for the present situation. The compact Lie group SU$(4)$ is generated by a set of 15 generalized Gell-Mann matrices $\{\hat{\lambda}_1,...,\hat{\lambda}_{15}\}$, see App.~\ref{GellMaaan} for an explicit representation. The initial interaction, given by the contributions $\sim J_{1,\Lambda}$ in Tab.~\ref{InitWWStruc} is invariant under a subgroup, generated by seven of these,
\begin{equation}
\label{GenJ1}
\m{Gen}_{\m{SU}(4)} = \{\hat{\lambda}_1,\hat{\lambda}_6,\hat{\lambda}_7,\hat{\lambda}_{12},\hat{\lambda}_{13},\hat{\lambda}_{14}\}.
\end{equation}
All of these generators may in principle be subject to spontaneous breaking. Consider now a homogeneous ansatz for the order parameter, i.e. the one corresponding to the BZA phase, in the four-sublattice description,
\begin{equation}
\label{BZA_Q}
\hat{Q}_{\Lambda\rightarrow\Lambda_{\rm{UV}}}^{\rm{BZA}} = Q_{\rm{UV}}\cdot\begin{pmatrix} 0 & \mathbbm{1}_2 +\sigma_x\\ \mathbbm{1}_2+\sigma_x & 0 \end{pmatrix}.
\end{equation}
This ansatz does not commute with any of the generators given in Eq.~\eqref{GenJ1}, separately. However, the combination
\begin{equation}
\hat{\lambda}_\pi = \hat{\lambda}_1 + \hat{\lambda}_6\,,
\end{equation}
leaves the interaction as well as the order parameter, Eq.~\eqref{BZA_Q}, invariant. It thus generates an unbroken subgroup.

Indeed, when implementing the special configuration, Eq.~\eqref{BZA_Q}, corresponding to $Q^{(1)} = Q^{(2)} = Q^{(3)} = Q^{(4)}$, as an initial condition for the RG flow equations, one finds that the divergence is \emph{not} regularized and solving the flow all the way through $\Lambda\rightarrow 0$ is not possible. The conclusion is that the physics of the $J_1$ Heisenberg model requires the subgroup generated by $\lambda_\pi$ to be broken as well. To find an appropriate order parameter, all one has to do is to find generators of SU(4) that do not commute with $\lambda_\pi$ and add at least one of them to the ansatz from Eq.~\eqref{BZA_Q}.
Eligible candidates are the Gell-Mann matrices
\begin{subequations}
\label{FluxGens}
\begin{align}
\m{Gen}_{\m{flux},\mathbbm{R}} &= \{\hat{\lambda}_2,\hat{\lambda}_3,\hat{\lambda}_4,\hat{\lambda}_5\} \quad \mbox{ and }\\
\m{Gen}_{\m{flux},\mathbbm{I}} &= \{\hat{\lambda}_8,\hat{\lambda}_9,\hat{\lambda}_{10},\hat{\lambda}_{11}\}.
\end{align}
\end{subequations} 
Not surprisingly, they correspond to modifications of the real or imaginary parts of the initial values of $Q^{(1..4)}$, respectively. Thus, each ansatz with expansion coefficients $Q_i\in \mathbbm{R}$
\begin{equation}
\label{FluxAnsatz}
\hat{Q}_{\Lambda_{\rm{UV}}}^{\rm{BZA}} + \sum_i Q_i\hat{\lambda}_i,\ \hat{\lambda}_i \in \left\{\m{Gen}_{\m{flux},\mathbbm{R}}\cup\m{Gen}_{\m{flux},\mathbbm{I}}\right\},
\end{equation}
leads to a maximally broken phase where the RG flow can be expected to converge and (almost) any random ansatz generates a valid flow in the infrared, as observed in Sec.~\ref{Sub1_OP} .

In order to understand the physical meaning of the generators, Eq.~\eqref{FluxGens}, recall that the initial interaction, Eq.~\eqref{BaseHamil}, in the pseudofermion formulation is invariant under an artificial local U(1) symmetry. 
One finds that the generators $\hat{\lambda}_i$ inside of the fermionic order parameter structure
\begin{equation}
\Psi^\dagger \hat{\lambda}_i\Psi, \quad \hat{\lambda}_i \in \left\{\m{Gen}_{\m{flux},\mathbbm{R}}\cup\m{Gen}_{\m{flux},\mathbbm{I}}\right\}\,,
\end{equation}
can be converted into each other by such gauge transformations. 
The notion of \textit{translational symmetry breaking} in this context should therefore always be understood modulo the application of such local transformations~\cite{Affleck1988}.
%, which is compatible with the literature on the subject

From the algebraic considerations alone, it is not obvious why the very general ansatz, Eq.~\eqref{FluxAnsatz}, should always lead to a $\pi$-flux phase in the infrared. This must indeed be shown by actually solving the flow equations. However, two important conclusions can be drawn already: Firstly, the sublattice ansatz allows to detect finite-ranged spatial symmetry breaking patterns by the familiar divergences. The range can be extended systematically by using larger numbers of sublattices. Secondly, the (non-)occurrence of divergences for specific symmetry breaking or preserving ans\"atze conveys information about which state is energetically preferred and thus closer to the actual ground state of the system. It thus makes a direct comparison of free energies unnecessary. 
To emphasize this key feature of our FRG approach, we reiterate that solving the flow equations therefore automatically selects the symmetry-breaking pattern and identifies the energetically favored ground state.
The reliability of our approach strongly benefits from this particular finding, as our present truncation scheme is not specifically tailored to compute precise values of the free energy.

%------------------------------------
\subsection{Frustration at large $N$}
\label{MFResults}
%------------------------------------

\begin{figure}[t!]
\includegraphics[width=\columnwidth]{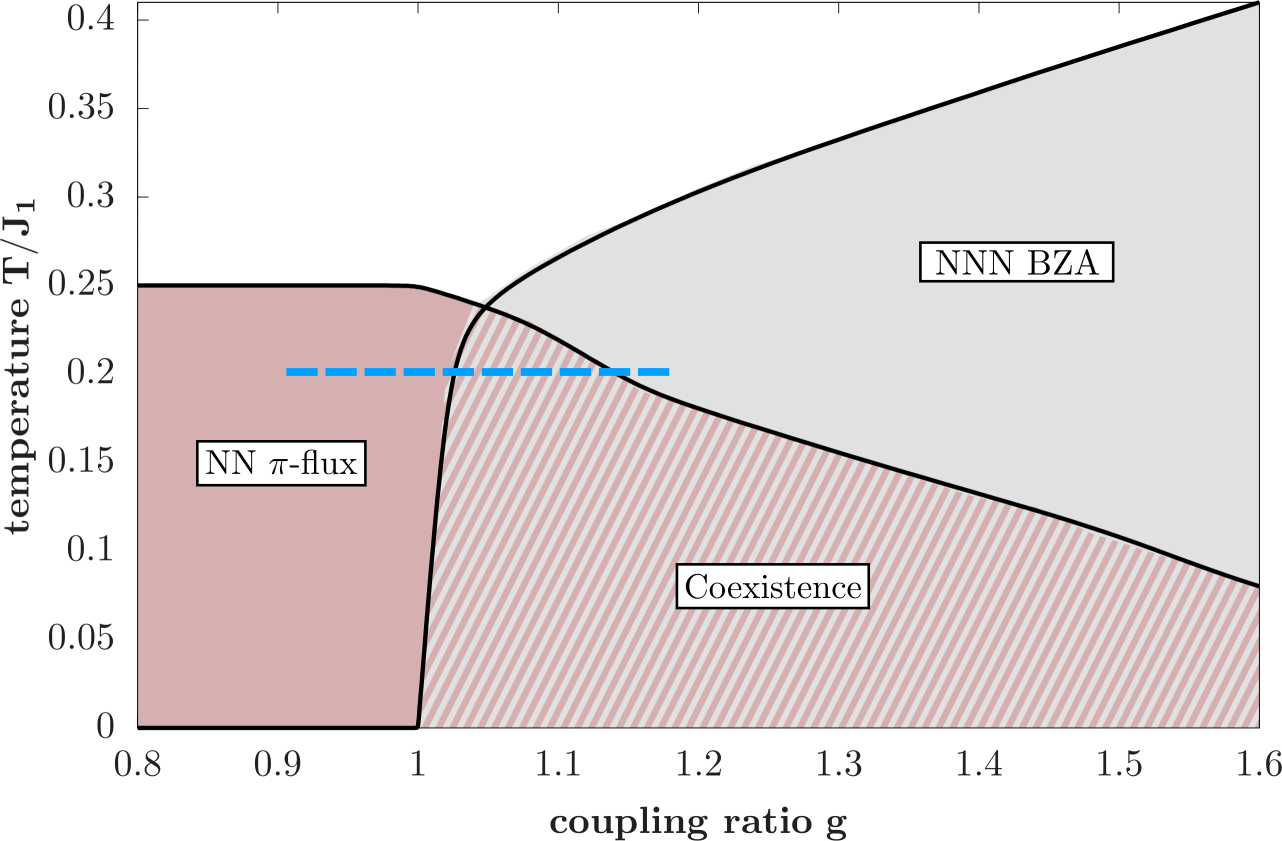}
\caption{Finite-temperature phase diagram of the $J_1$-$J_2$-model at large $N$. A nearest-neighbor $\pi$-flux phase (dark shading) is found for all temperatures and $g<1$. For $g>1$, next-to-nearest neighbor order parameters become finite, while there is an extended coexistence phase (striped shading) at low temperatures, separated from the high-$T$ normal phase by a next-to-nearest neighbor BZA phase. The blue (dashed) line designates the parameter range of Fig.~\ref{CoexistSlice}.}
\label{NPhaseDiag}
\end{figure}

We now include magnetic frustration and consider the $J_1$-$J_2$ model for a broad range of the ratio $g = \frac{J_2}{J_1}$. To that end, we take into account the full initial action and order parameter structure given in Tabs.~\ref{InitWWStruc} and~\ref{InitQStruc}. The number of couplings generated during the flow increases, too, summing up to 48 differential equations to be solved. The resulting finite-temperature phase diagram is shown in Fig.~\ref{NPhaseDiag}. 
For sufficiently low temperature, an ordered phase is found for all $g$. While $Q^{(1..4)}$ dominate for $g<1$, finite $Q^{(5,6)}$ become mandatory to describe the phase at $g>1$. The former exhibits $\pi$-flux characteristics, while $Q^{(5,6)}$ do not entail a complex phase for next-nearest neighbor plaquettes. This particular finding is most likely due to our limited spatial range: with four sublattices, no next-to-nearest neighbor $\pi$-flux phase can be described. We would expect to find such a flux phase for $g>1$ as well, provided the number of sublattices would be sufficiently increased. 
Since the critical temperature does not depend on the spatial structure of the order parameter in this system~\cite{AffleckMarston88,ArovasAuerbach88}, we can still expect to obtain qualitatively correct results.

%------------------------------------
%------------------------------------

\begin{figure}[t!]
\includegraphics[width=\columnwidth]{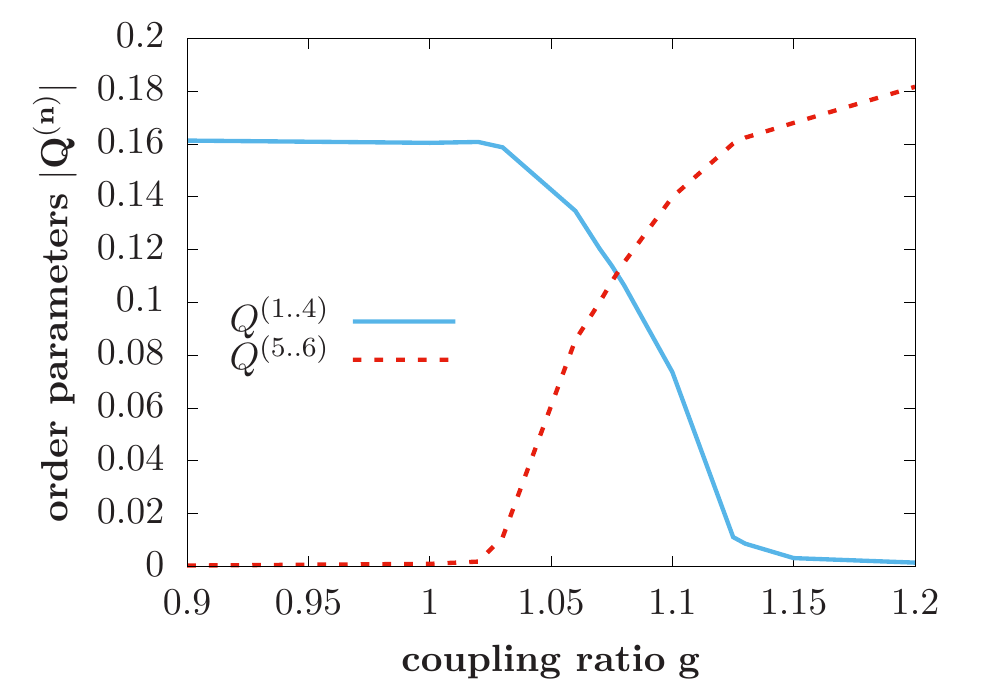}
\caption{Absolute values of nearest (solid line) and next-to nearest (dotted line) neighbor order parameters for $T=0.2$ in the coexistence region ($Q^{(1..4)}\neq 0$ and  $Q^{(5,6)} \neq 0$), see Fig.~\ref{NPhaseDiag}.}
\label{CoexistSlice}
\end{figure}

For $ g>1$ and low temperatures, we find a regime of coexistence and a competition of ordering tendencies is indicated, cf. Figs.~\ref{NPhaseDiag} and~\ref{CoexistSlice}, by the suppression of the critical temperature or the order parameter in the coexistence region.
Indeed, frustration is expected to occur due to a finite $J_2$. However, it does not seem sensible to directly relate this finding to the suspected spin liquid regime of the physical model at $N=2$: Firstly, it is widely agreed~\cite{Chandra1988,ED95,ED96,ED00,SpinWave06,ED06,ED09,MasterEq09,DMRG12,Variational12,Variational13,ClusterMF13,DMRG13,RG14,CC14,MC14,TPS16,PlaqExp16,DMRG17,PEPS17,PEPS18,LatestPEPS18}
that the latter is situated at about $0.4\lesssim g\lesssim 0.6$ which does not even overlap with the large-$N$ coexistence region. Secondly, it is doubtful whether \textit{frustration} is an appropriate term to describe the mechanism behind this coexistence. At $N\rightarrow\infty$, substantial contributions of the interactions are suppressed~\cite{LargeNReal}, notably those being responsible for magnetic ordering tendencies~\cite{BaezLargeS17}. Those very interactions, commonly referred to as large-$S$ enhanced, are known to facilitate frustration~\cite{Chandra1988}. In our case, instead of being cancelled by (geometrically) competing spin-spin interactions, magnetic ordering is thus suppressed artificially, giving way to a competition between \emph{spin-liquid} phases instead.

If the mechanism behind the putative spin-liquid phase at $N=2$ is indeed a suppression of magnetic ordering due to frustration and large-$N$ results become a good approximation in this regime, we would therefore expect a nearest-neighbor dominated spin-liquid phase like a $\pi$-flux phase to occur.

%------------------------------------
%------------------------------------

%------------------------------------
\section{Phase diagram at $N=2$}
\label{Ngleich2}
%------------------------------------

At $N=2$, additional fluctuations occur which are absent in the large-$N$ limit. After discussing the implementation of the fermion number constraint, we therefore extend the truncation by taking into account further order parameter structures and newly generated four-fermion interactions. 
Then, we calculate the phase diagram of the $J_1$-$J_2$ model for $g\in [0,1]$ with our approach and exhibit the suppression of magnetic correlations upon inclusion of a minimal frequency-dependent self-energy $\Gamma^{(\gamma)}$.
In fact, we show how $\Gamma^{(\gamma)}$ leads to a phase diagram which is consistent with the results from previous pf-FRG studies.
To improve the understanding of the role of $\Gamma^{(\gamma)}$, we briefly revisit the large-$N$ case and show that, here, the spin-liquid state distinctly originates from the flow of the four-fermion vertex and not from a frequency-dependent self-energy. Including an artificial $\Gamma^{(\gamma)}$ at large $N$ leads to a strong suppression of the spin-liquid state, too.
Based on this consideration, we return to $N=2$ and argue that the inclusion of frequency-dependent self-energy induces a damping of both, magnetic as well as bilinearly-ordered spin-liquid states, and therefore does not support the formation of the latter.

%------------------------------------
\subsection{Fermion number constraint}
\label{PopotovRulez}
%------------------------------------

For large $N$, it is sufficient to implement the fermion number constraint~\eqref{OpConstraint} on average~\cite{ArovasAuerbach88}, which greatly simplifies the analysis. Moving to $N=2$, the problem has to be dealt with more carefully. There are at least two viable options on how to faithfully implement the constraint in this situation.

The first and probably better known one is to make use of an artificial local SU(2) symmetry of the model, Eq.~\eqref{BaseHamil}, introduced due to the pseudofermion formulation~\cite{Affleck1988}. The constraint as expressed in Eq.~\eqref{OpConstraint} explicitly breaks this symmetry down to a local U(1). 
However, implementing the constraint in the form of a Lagrange multiplier field $\vec{A}$ restores the local SU(2),see Ref.~\onlinecite{Affleck1988} for details. Moreover, the local SU(2) now becomes a true gauge symmetry as the Lagrange multiplier plays the role of a gauge field. The confinement physics associated with this gauge field is often taken as a hallmark of spin liquid behavior~\cite{Savary2017}.
Although the predictive power associated with the gauge field is persuasive, it entails technical challenges that are beyond the scope of this work~\footnote{ 
Dealing with (non-abelian) gauge fields in general is not a easy task within the FRG formalism. It could at best be done in an approximate way and it is not clear in which way a given approximation scheme would influence the fulfillment of the constraint and thus the physics outcome. Furthermore, the field $\vec{A}$ is expected to acquire its own dynamics during the flow. Since we are considering a lattice system, such a gauge field is generally compact and a proper description might have to include vortex or monopole degrees of freedom. Although there is work in this direction in the context of BKT transitions~\cite{Machado2010,Krieg2017}, this represents a formidable task on its own and is therefore beyond the scope of this paper.}. Instead, we follow a different approach, here.

The other realization of the constraint was put forward by Popov and Fedotov~\cite{PopovFedotov88}, who showed that equipping the pseudo-fermion Hamiltonian with an imaginary chemical potential
\begin{equation}
\label{PopotovDef}
\mu_{\rm{PF}} = i \frac{\pi T}{2}\,,
\end{equation}
leads to an exact cancellation of all unphysical contributions in the partition function. 
The constraint is thus implemented exactly on the level of the microscopic action~\eqref{BareS}. This procedure is suitable for our FRG approach and the ansatz for the bilinear part of the effective average action $\Gamma_\Lambda^{(2\Psi)}$ without order parameters becomes
\begin{equation}
\label{PopotovinS}
\Gamma_\Lambda^{(2\Psi)} = \sum_n\int_{p\in \m{TZ}}\left[ \Psi_p^\dagger \left(i\omega_n\vartheta_\Lambda^{-1} + \mu_\Lambda\right)\mathbbm{1}\Psi_p\right],
\end{equation}
with $\mu_{\Lambda_{\rm{UV}}} = \mu_{\m{PF}}$. Here, $\vartheta_\Lambda^{-1}$ is the formal inverse of the multiplicative regulator function~\cite{EnssThesis}
\begin{equation}
\label{EnssReg}
\vartheta_\Lambda = \left\{ \begin{matrix} 0 & |\omega_n| < \Lambda-\pi T \\ \frac{1}{2} + \frac{|\omega_n|-\Lambda}{2\pi T} & \Lambda-\pi T \leq |\omega_n| \leq \Lambda + \pi T\\ 1 & \Lambda + \pi T \leq |\omega_n| \end{matrix}\right..
\end{equation}
There is a number of things to be observed about this way of implementing the constraint. Firstly, it is not equivalent to a functional $\delta$~function enforcing Eq.~\eqref{OpConstraint} as in the case of the Lagrange multiplier. Hence, it is not permissible to simply replace any $f^\dagger_{i\alpha} f^\nodagger_{i\alpha}$ with the number 1 in the action. 
Consequently, the structure of the initial interaction is now more complicated than the one given in Eq.~\eqref{BareS}, because 
\begin{equation}
\label{DensExtra}
S^\mu_i S^\mu_j = -\frac{1}{2}f^\dagger_{i\alpha} f^\nodagger_{j\alpha}f^\dagger_{j\beta} f^\nodagger_{i\beta} - \frac{1}{4}f^\dagger_{i\alpha} f^\nodagger_{i\alpha}f^\dagger_{j\beta} f^\nodagger_{j\beta}\,,
\end{equation}
and the last term on the right-hand side, corresponding to a density-density interaction, cannot be neglected anymore.

Secondly, the imaginary chemical potential in the ansatz Eq.~\eqref{PopotovinS} is equipped with an explicit $\Lambda$~dependence, its initial value set to $\mu_{\m{PF}}$. This may seem odd at first glance, since the constraint~\eqref{OpConstraint} is an exact relation that does not seem to lend itself to a continuous RG flow. One should note, however, that the proof of $\mu_{\m{PF}}$ implementing the constraint was given for the microscopic Hamiltonian, i.e. at the scale $\Lambda = \Lambda_{\m{UV}}$. The value of the partition function does, of course, not change under an (exact) RG transformation, but the shape of the Hamiltonian or the action does. Physically, this means that the $f^{(\dagger)}$ fields are systematically dressed. To fulfill the constraint at scales $\Lambda\neq \Lambda_{\m{UV}}$, it is therefore mandatory to let $\mu_\Lambda$ evolve with the rest of the effective average action.

\begin{figure}[t!]
\includegraphics[width=\columnwidth]{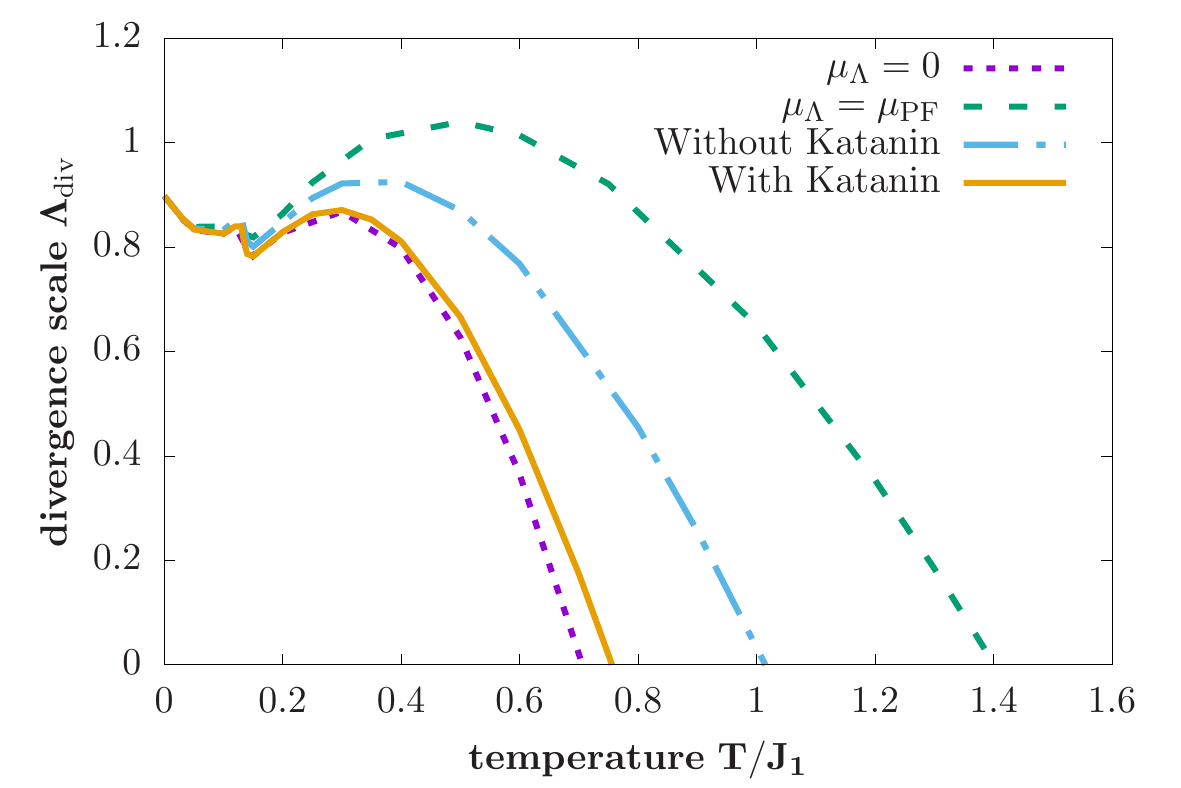}
\caption{Divergence scale $\Lambda_{\m{div}}$ as function of temperature for different approximation schemes of the toy model in Tab.~\ref{ToyInts}. See main text for details.}
\label{PopotovSSB}
\end{figure}

Our approach to the solution of the FRG equation~\eqref{Wettereq} is approximate and so is the renormalization of $\mu_{\Lambda}$. Therefore, the implementation of the constraint is not exact for any $\Lambda < \Lambda_{\m{UV}}$, just as for a hypothetical gauge field realization. On the other hand, we do have better control about the approximation error since we keep the truncation exactly at the same level as for the rest of the action.
Let us illustrate this point with a toy model of our system. Consider $N=2$ in the limit $J_2=0$ without initial order parameters on two sublattices, i.e. $\Psi$ is a two-dimensional spinor. The latter is necessary for an algebraic discrimination between the two contributions to the right hand side of Eq.~\eqref{DensExtra} in momentum space. The four-fermion terms that are present initially or will be generated during the flow are shown in Tab.~\ref{ToyInts}.
\begin{table}[H]
\centering
\begin{tabular}{|c|c|c|c|}
\hline
 & & & \\[-1em]
Coupling & $\eta^X$ & $\eta^Y$ & $f(p_1,p_2,p_3,p_4)$\\
\hline
$J_{\Lambda}$ & $\sigma^+$ & $\sigma^-$ & $\cos(p_{2,x} - p_{3,x}) +\cos(p_{2,y} - p_{3,y})$\\
$L_{\Lambda}$ & $\sigma^o$ & $\sigma^u$ & $\cos(p_{3,x} - p_{4,x}) +\cos(p_{3,y} - p_{4,y})$\\
\hline
$X_\Lambda$ & $\sigma^o$ & $\sigma^o$ & 1 \\
$X_\Lambda$ & $\sigma^u$ & $\sigma^u$ & 1 \\
\hline
\end{tabular}
\caption{Full interaction structure of the symmetric two-sublattice $J_1$ model. Matrix definitions are: $\sigma^{\pm} = \frac{1}{2}(\sigma^x\pm\sigma^y)$ and $\sigma^{ou} = \frac{1}{2}(\mathbbm{1}\pm\sigma^z)$ with the usual definitions of the Pauli matrices $\sigma^{x/y/z}$.}
\label{ToyInts}
\end{table}
The flow equations for the couplings $J_\Lambda$, $L_\Lambda$, $X_\Lambda$ and $\mu_\Lambda$ can be found in App.~\ref{ToyEquations}. Depending on the temperature, the flow is intercepted by a divergence at some scale $\Lambda_{\m{div}}$. While this scale itself does not bear direct physical meaning, its existence signals the onset of ordering~\cite{Braun2012}. The highest temperature where $\Lambda_{\m{div}} > 0 $ will be dubbed \textit{instability temperature}.
In Fig.~\ref{PopotovSSB}, the value of $\Lambda_{\m{div}}$ is plotted for a number of different approximation schemes: no constraint ($\mu_\Lambda = 0$, dotted line), static chemical potential ($\mu_\Lambda = \mu_{\m{PF}}$, dashed line), running chemical potential with (solid line) or without (dot-dashed line) Katanin-feedback of $\mu_\Lambda$ on the four-fermion couplings. The impact of these approximations on the instability temperature is quite dramatic and it is therefore vital to include $\mu_\Lambda$ into a consistent approximation scheme~\footnote{In previous work~\cite{LargeNMomentum} we suggested a scheme that would correspond to the case of a static $\mu_\Lambda$. In light of the present findings, we now rather advocate the complete Katanin scheme (solid line in Fig.~\ref{PopotovSSB}) as the latter is used for the flow with order parameters as well.}.

One last remark is due to the zero-temperature limit of the problem. By definition in Eq.~\eqref{PopotovDef}, $\mu_{\m{PF}}$ vanishes for $T\rightarrow 0$. It is not obvious, whether this limit commutes with the path integral, i.e. the computation of the partition function. At least in our toy model, this seems to be the case, though. For all approximations, $\Lambda_{\m{div}}$ smoothly approaches the na\"ive zero-temperature value.

%------------------------------------
\subsection{Extended truncation}
\label{Pairing}
%------------------------------------

In the frustrated regime of the $N=2$ case, it is insufficient to consider only the order parameters $Q^{(1..4)}$ and $Q^{(5,6)}$ due to the presence of additional fluctuations away from the large-$N$ limit.
Therefore, we have to introduce additional pairing order parameters $\Delta$, allowing us to investigate more classes of possible spin-liquid phases. Our ansatz for the bilinear part $\Gamma_\Lambda^{(2\Psi)}$ of the effective average action  becomes
\begin{equation}
\label{GBilinear}
\begin{aligned}
\Gamma_\Lambda^{(2\Psi)} =& \sum_{n,\alpha\beta}\int_{p\in TZ}\Big[ \Psi_{p\alpha}^\dagger \left(i\omega_n\vartheta_\Lambda^{-1} + \mu_\Lambda\right)\mathbbm{1}\delta_{\alpha\beta}\Psi_{p\beta} \\
&+ \sum_{i=1}^6\Psi_{p\alpha}^\dagger\left(Q_\Lambda^{(i)}\eta_i + Q_\Lambda^{(i),*}\eta^{T}_i\right)\delta_{\alpha\beta}\mathfrak{Q}(p)\Psi_{p\beta}\\
&+\sum_{j=1}^6 \Delta^{(j),*}\Psi^\dagger_{p\alpha}\eta^{T}_j\Psi^*_{-p\beta}\epsilon_{\alpha\beta}\mathfrak{D}(p)\\
&+ \sum_{j=1}^6 \Delta^{(j)}\Psi_{p\alpha}^T\eta_{j}\Psi_{-p\beta}\epsilon_{\alpha\beta}\mathfrak{D}(p) \Big]\,.
\end{aligned}
\end{equation}
Here, $\mathfrak{Q}(p)$ and $\mathfrak{D}(p)$ are the respective geometric momentum structures. The pairing order parameters are associated with a non-trivial structure $\sim \epsilon_{\alpha\beta}$ in spin space preserving the global SU(2) symmetry. Instead of the totally antisymmetric tensor $\epsilon_{\alpha\beta}$, Pauli matrices $\sigma^{x,z}_{\alpha\beta}$ or linear combinations would be permissible as well. Due to SU(2) symmetry, however, this does not change the structure of the flow equations.

Introducing the new order parameters has consequences for the four-fermion sector of the model as well. For $\Delta^{(j)}=0$, Eq.~\eqref{SubWWStruc} is the only possible interaction structure, different combinations of $\eta^{X/Y}$ determining the number of couplings to be taken into account. Once any $\Delta^{(j)} >0$, there are five new structures that can and generally will be generated during the flow, see Tab.~\ref{NewPInts}. Note that, SU(2) symmetry keeps the spin structure of the interactions simple in the sense that $\delta_{\alpha\beta}$ and $\epsilon_{\alpha\beta}$ are associated to fixed Nambu sectors during the entire RG flow.
\begin{table}[H]
\centering
\begin{tabular}{|c|c|c|}
\hline
Type & Grassmann Structure & Momentum Structure\\
\hline
\hline
 & & \\[-1em]
Original & $(\Psi^\dagger\eta^X\delta\Psi)(\Psi^\dagger\eta^Y\delta\Psi)$ & $f_k(1,-1,1)$\\
\hline
 & & \\[-1em]
\multirow{2}{*}{Mixed} & $(\Psi^\dagger\eta^X\epsilon\Psi^*)(\Psi^\dagger\eta^Y\delta\Psi)$ & $f_k(-1,-1,1)$\\
 & $(\Psi^T\eta^X\epsilon\Psi)(\Psi^\dagger\eta^Y\delta\Psi)$ & $f_k(1,-1,1)$\\
\hline
 & & \\[-1em]
\multirow{3}{*}{Pairing} & $(\Psi^\dagger\eta^X\epsilon\Psi^*)(\Psi^\dagger\eta^Y\epsilon\Psi^*)$ & $f_k(1,1,1)$\\
 & $(\Psi^T\eta^X\epsilon\Psi)(\Psi^T\eta^Y\epsilon\Psi)$ & $f_k(1,1,1)$\\
 & $(\Psi^\dagger\eta^X\epsilon\Psi^*)(\Psi^T\eta^Y\epsilon\Psi)$ & $f_k(1,-1,-1)$\\
\hline
\end{tabular}
\caption{New interaction structures due to finite pairing order parameters. Spin space is represented by the diagonal matrix $\delta$ and the totally antisymmetric matrix $\epsilon$. The geometric structure $f_k$ is given in Eq.~\eqref{NNMomStruc}.}
\label{NewPInts}
\end{table}
As opposed to the large-$N$ case, all combinations of $\eta^{X/Y}$ matrices can now contribute to the RG flow. Taking into account all of these combinations for all Grassmann structures would correspond to a total of 1536 couplings in the four-fermion sector. There is, however, a number of symmetries that can be exploited: for instance, the two mixed Grassmann structures are hermitean conjugates of each other and thus do not contain independent information. Furthermore, the choice of the $\eta^{X/Y}$ matrices themselves is constrained by hermiticity as well. Taking all of this into account reduces the number of independent (complex) couplings from the four-fermion sector down to 624. The flow equations for these are then solved by step-size-controlled numerical integration.

%------------------------------------
%------------------------------------

When translating the initial action~\eqref{BareS} into momentum space, the definition of nearest or next-nearest neighbor coupling guarantees a unique representation of the vertex' momentum structure with respect to the choice of sublattices. In contrast to the large-$N$ case, this structure is not invariant under the RG flow at $N=2$ anymore. Longer-ranged interactions are generated. Whenever the range of such an interaction exceeds the enlarged unit cell given by the sublattice representation, the momentum structure is altered. In order to properly account for this effect, projection onto the momentum structure itself
would be required. 
Here, we strictly truncate our interaction vertices to take into account the nearest-possible neighbor sites, only. In practice, this corresponds to the following generic momentum structure
\begin{equation}
\label{NNMomStruc}
\begin{aligned}
f_k(i,j,l) = &\cos(ia_xk_2 + jb_xk_3 + lc_xk_4)\\
&\cdot\cos(ia_yk_2 + jb_yk_3 + lc_yk_4)\,.
\end{aligned}
\end{equation}
The $k_i$ are the momenta of the vertex' fields. $\{a_{x/y},b_{x/y},c_{x/y}\}$ are determined by $\eta^{X/Y}$, i.e. the geometry of the vertex and $\{i,j,k\}$ vary with the Grassmann sector considered, see Tab.~\ref{NewPInts}. For $N\rightarrow\infty$, we recover the momentum structures employed in Sec.~\ref{Sub1_OP}.

We do not expect this truncation to interfere with the regularization of the RG flow through emergent spin-liquid order. This is in analogy to the analysis presented in Sec.~\ref{Sub2_FancyGroups}: there, we showed that, in the two-sublattice parametrization, the uniform BZA state regularizes the flow, but in the four-sublattice parametrization the additional translational symmetry breaking of the $\pi$-flux state is exhibited. The level of truncation was equivalent to eq.~\eqref{NNMomStruc} in both instances. We expect this mechanism to extend to parametrizations with more sublattices.

%------------------------------------
\subsection{Phase diagram \& frequency dependence}
\label{FreqDiss}
%------------------------------------

We first discuss our numerical findings for $J_2=0$. In this case, a N\'{e}el-ordered magnetic phase is expected to occur at sufficiently low temperatures~\cite{ManousakisRev91}. To compute infrared properties of such a phase in our approach, one would have to introduce magnetic order parameters, i.e. terms $\sim M^{(n)} \Psi^\dagger_\alpha \tau_{\alpha\beta}\eta^X\Psi_\beta$ where $\tau_{\alpha\beta}\neq \delta_{\alpha\beta}$ in spin space. Here, we are not interested in magnetic orders and we refrain from taking them into account. This greatly reduces the numerical cost.
For $T/J_1<0.82$ we find a divergence of the four-fermion couplings that does not trigger a commensurate growth of any of the viable bilinear spin-singlet orders $Q^{(i)}_\Lambda$ or $\Delta^{(j)}_\Lambda$. Therefore, the divergence indicates a spontaneous breaking of the global SU(2) and hence magnetic ordering. 

\begin{figure}[t]
\includegraphics[width=\columnwidth]{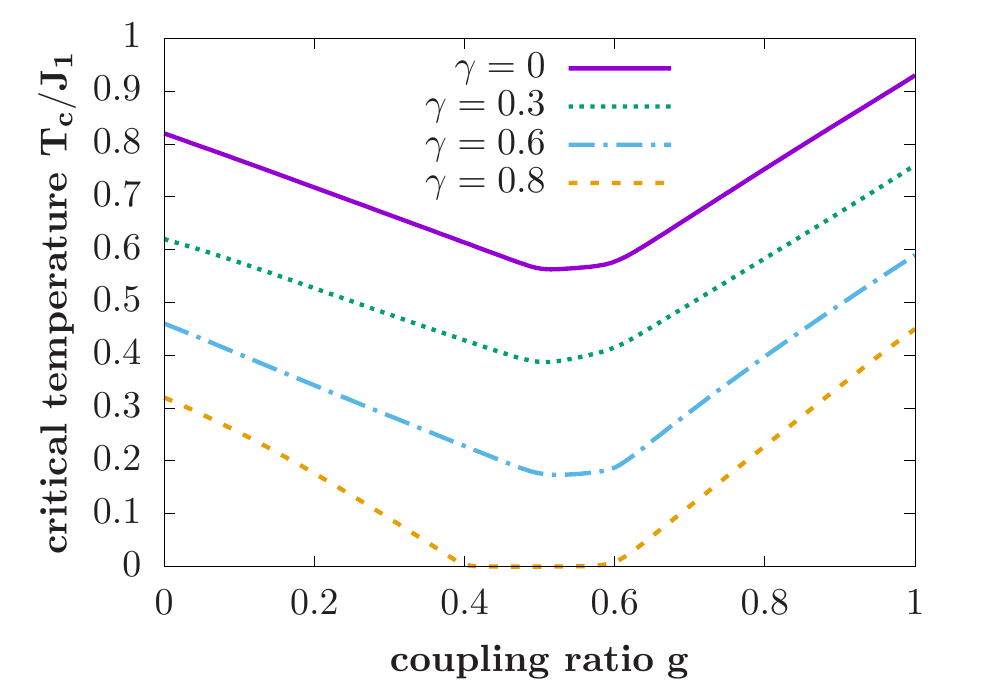}
\caption{Critical temperatures $T_c/J_1$ for magnetic ordering for different values of the artificial damping constant $\gamma$ at $N=2$.}
\label{N2gammaPD}
\end{figure}

Considering $J_2>0$, the phase diagram in Fig.~\eqref{N2gammaPD} ensues (solid line), where magnetic ordering occurs for all $g=J_2/J_1$ with just a little reduction of the critical temperature around the suspected regime of frustration, $g\sim 0.5$.
This gross overestimation of magnetic ordering tendencies is expected in the presented \emph{static} approximation, where explicit frequency dependences of the four-fermion vertices or self-energy contributions have been neglected:
properly taking frequency dependence into account within the pf-FRG scheme~\cite{Reuther2010} leads to a magnetic phase diagram which is well compatible with the bulk of the literature, in particular with regard to the existence of a non-magnetized regime for $0.4\lesssim g\lesssim 0.6$.

Here, we shed light on the mechanism behind the suppression of magnetic ordering aiming at a better characterization of the spin liquid behavior that is facilitated instead:
in Ref.~\onlinecite{Reuther2010} it was shown that the phenomenological consequences of frequency dependence, such as finite pseudofermion lifetime, can be reproduced within the static approximation if an ad-hoc self energy
\begin{equation}
\label{gammaDef}
\Gamma_\Lambda^{(\gamma)} = \sum_n\int_{p\in TZ} i\gamma\Psi^\dagger_p \m{sign}(\omega_n)\mathbbm{1}\Psi^\nodagger_p %\,,
\end{equation}
is added to the ansatz for the effective average action. Being associated to an odd frequency structure, $\gamma$ is not renormalized as long as the vertex functions remain frequency-independent~\cite{Reuther2010}. By choosing larger values for $\gamma$, magnetic order is more and more suppressed until, eventually, the regime around $g\sim 0.5$ is indeed free of magnetic instabilities %phases
all the way down to $T=0$, cf. Fig.~\ref{N2gammaPD}.

%------------------------------------
%------------------------------------

%------------------------------------
\subsection{Large-$N$ spin liquid \& frequency dependence}
\label{FreqDissLargeN}
%------------------------------------

\begin{figure}[t]
\includegraphics[width=0.9\columnwidth]{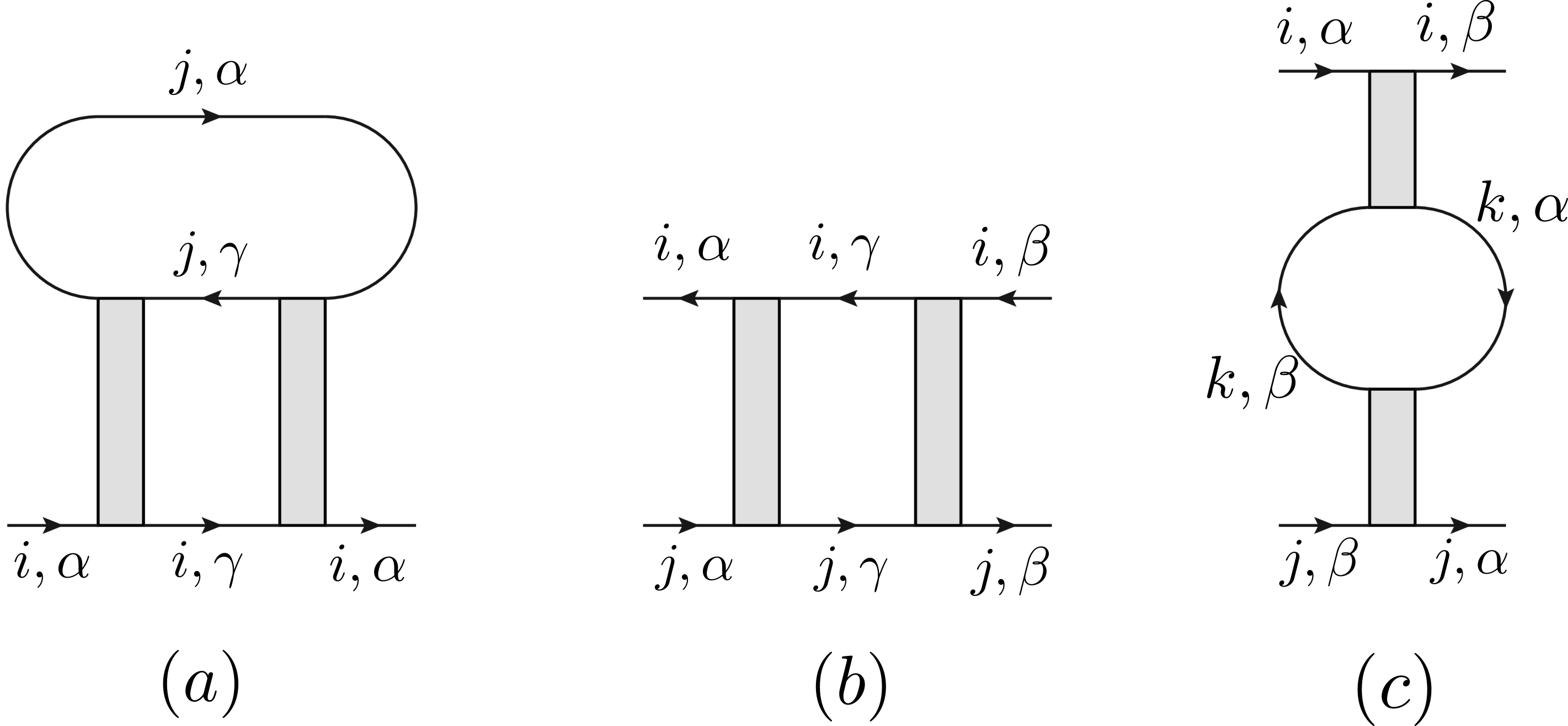}
\caption{Diagrammatic contributions to the RG flow. $\{i,j,k\}$ are site-indices and $\{\alpha,\beta,\gamma\}$ are general SU($N$) spin indices. (a) Two-loop self-energy diagram responsible for the imaginary contribution in Eq.~\eqref{gammaDef}. Its contribution is suppressed for $N\rightarrow\infty$. (b) Particle-hole contribution to the interaction vertices, dominant at large $N$. (c)~RPA-type contribution to the interaction vertices.}
\label{Sunrise}
\end{figure}

To assess the influence of frequency dependence on possible spin liquid phases, let us return for a moment to the large-$N$ situation: There, no change of the critical temperature is found when comparing a fully frequency dependent pf-FRG scheme to the simplified one at hand~\cite{LargeNReal,LargeNMomentum}. The reason for this behavior can be understood from perturbative diagrammatics. An imaginary self-energy like~\eqref{gammaDef} can naturally not occur from a one-loop diagram as the initial value for the four-fermion vertex is frequency independent. 
At two loop order, the diagram given in Fig.~\ref{Sunrise}a accounts for a non-trivial renormalization of the frequency dependence of the self-energy. In our FRG flow, it is effectively generated by a dressed vertex that is fed into the flow of the self energy. Since only one of the two loops contributes a factor of $N$ due to a freely summed spin index, the overall value of the diagram is $\sim\frac{1}{N}$ and thus suppressed for $N\rightarrow\infty$. Thus, even if the frequency dependence of vertices was taken into account, no such dependence of the self energy can be generated at large $N$. It is therefore safe to say that in the present system, the occurrence of spin-liquid phases at large $N$ is solely due to the peculiar structure of the four-fermion vertex flow: only the diagram with a particle-hole loop (Fig.~\ref{Sunrise}b) contributes while, in particular, the RPA-type diagram (Fig.~\ref{Sunrise}c), which is typically associated with magnetic ordering, is suppressed.

%------------------------------------
%------------------------------------

%------------------------------------
\subsection{$N=2$ case}
\label{FreqDissN2}
%------------------------------------

\begin{figure}[t]
\includegraphics[width=\columnwidth]{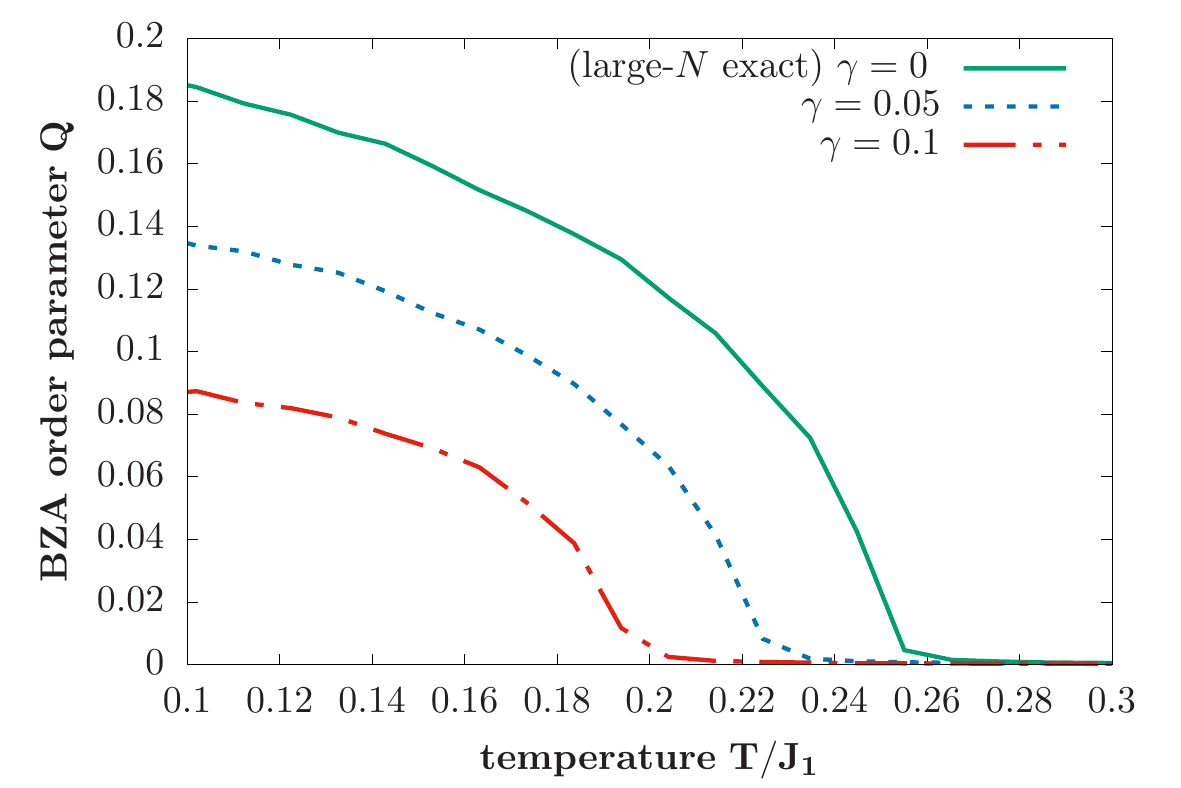}
\caption{Large-$N$ BZA phase order parameter for different values of the artificial damping $\gamma$.}
\label{NgammaArt}
\end{figure}

Considering the case $N=2$, both of these two-particle diagram types \emph{and} the self energy~\eqref{gammaDef} become important. We do know the effect of finite damping $\gamma>0$ on magnetic ordering (see Fig.~\ref{N2gammaPD}), but its impact on potential spin liquid phases is still unclear, yet of utmost importance for the understanding of the highly frustrated $g\sim 0.5$ regime. To shed light on this question, we need to investigate the interplay of $\Gamma_\Lambda^{(\gamma)}$ and spin-liquid favoring fluctuations represented by the diagram in Fig.~\ref{Sunrise}b. To this end, we artificially introduce the self energy~\eqref{gammaDef} into the large-$N$ flow equations of the $J_1$ model for two sublattices, which are by construction dominated by this type of fluctuations. This gives us a handle on the influence of $\gamma>0$ on spin liquid ordering tendencies also beyond the large-$N$ limit. 
The so-obtained results are shown in Fig.~\ref{NgammaArt}.

We find that bilinear spin liquid ordering is suppressed upon introduction of an artificial damping $\gamma$ as well. In fact, at a value of $\gamma = 0.3$, no such phase could be discerned anymore. This is particularly noteworthy, since for $\gamma=0.3$, the suspected spin liquid regime around $g\sim 0.5$ at $N=2$ (cf. Fig.~\ref{N2gammaPD}, dotted line) is still occupied by magnetic phases. Running the complete $N=2$ flow equations again results in a divergence, no matter if the order parameters from Eq.~\eqref{GBilinear} are present or not. 

We draw two conclusions from these results. Firstly, frequency dependence alters quantitative features such as the values of critical temperatures or the position of phase boundaries. However, since damping as its main consequence affects both, conventional (magnetic) as well as bilinear spin liquid ordering, we do not expect the appearance of physical spin liquid phases which are not present in the static FRG approximation. This solidifies the reliability of our approach.
Secondly, the primary goal of this work was to shed some further light on the nature of the frustrated regime around $g \sim 0.5$. 
Importantly, we can conclude that our approach does not support the emergence of
any particular type of bilinearly ordered spin liquid, since we included all possible bilinear order parameters that do \emph{not} break the global SU(2) symmetry. 
This indicates that bilinear spin-liquid order is not an appropriate description of the physical phenomena occurring in the $J_1$-$J_2$ model in the regime $0.4\lesssim g \lesssim 0.6$. 

%------------------------------------
%------------------------------------

By construction, our approach is sensitive to the occurrence of any bilinear order parameter, signaled by a divergence in the four-fermion sector of the effective average action. However, ordering tendencies that need to be described by expectation values of more fermion fields are {\em not} seen at the given truncation level. The results presented, here, therefore strongly hint towards such a more complicated structure of the spin liquid phase. Possible candidates are dimer valence bond solids~\cite{SpinWave06,RG14,PEPS18} (involving four fermion fields) or a four-spin (eight-fermion) plaquette order~\cite{ED00,ED06,DMRG13,PlaqExp16}.

%------------------------------------
\section{Conclusions \& Outlook}
\label{Cloudlook}
%------------------------------------

To summarize, we have introduced a cluster pf-FRG approach and applied it to the $J_1$-$J_2$ Heisenberg model on the square lattice. First, a large-$N$ analysis indicated the necessity to incorporate spatially structured order parameters by means of the cluster extension. Moving to the physical case at $N=2$, we carefully analyzed the implementation of the fermion number constraint using an imaginary chemical potential according to the Popov-Fedotov method. We added a minimal frequency-dependent self-energy to our truncation to incorporate the effects of finite pseudofermion lifetime. By doing so we were able to reproduce the non-magnetic phase at intermediate coupling ratio $0.4\lesssim g \lesssim 0.6$. However, no signs of bilinear spin-liquid order emerged in this regime. In fact, any instability encountered in this model could reliably be related to magnetic ordering tendencies instead.
This suggests that the low temperature regime of the strongly frustrated $J_1$-$J_2$ Heisenberg model is occupied by a state that needs to be characterized by the expectation value of at least {\em four} fermion fields, the simplest incarnation of which includes valence bond solid or plaquette ordered states, but also more complex spin liquid states.

There are multiple directions for future work. As indicated in Sec.~\ref{Sub1_OP}, the sublattice representation opens up our method for further, more complicated lattice geometries. A natural next step is to perform an analogous analysis for the Kitaev model on the honeycomb lattice. While the exact solution is formulated in terms of Majorana fermions~\cite{Kitaev2006}, it has been shown that an equivalent pseudofermion representation exists~\cite{Burnell2011}. In this case, the QSL phase is indeed characterized by the order parameters introduced in Eq.~\eqref{GBilinear}. Having reproduced these findings, one should be able to analyze extensions of the Kitaev model with only minor additional effort.

%------------------------------------
\begin{acknowledgments}
We thank F. L. Buessen and D. Kiese for ongoing collaborations on related projects and comments on the manuscript. We acknowledge  valuable discussions  with J. Reuther and B. Sbierski and their comments on the manuscript. We also thank J. Braun and N. Dupuis for insightful discussions.
This work was partially supported by the Deutsche Forschungsgemeinschaft (DFG, German Research Foundation) -- Projektnummer 277146847 -- SFB 1238 (project C03).
S.D. and D.R. acknowledge support by the German Research Foundation (DFG) 
through the Institutional Strategy of the University of Cologne within the German Excellence Initiative (ZUK 81).
\end{acknowledgments}
%------------------------------------

%------------------------------------
\appendix
%------------------------------------

\section{Large-$N$ structure of the fluctuation matrix}
\label{NFlucMat}
The right hand side of the flow equation~\eqref{Wettereq} is determined by the second functional derivative of the effective average action, 
$
\Gamma_\Lambda^{(2)} = \mathcal{P}_\Lambda + \mathcal{F}_\Lambda.
$
Here, $\mathcal{F}_\Lambda$ is the field-dependent part of $\Gamma_\Lambda^{(2)}$, called fluctuation matrix. Applying the point-like projection $\Psi_{\alpha,\omega_m-\nu_n,\p{p-q}} = \Psi_\alpha \delta_{m,n}\delta(\p{p-q})$, its value with respect to the generic interaction~\eqref{SubWWStruc} is given by
\begin{equation}
\mathcal{F}_\Lambda = \frac{J_n}{N}\delta_{m,n}\delta(\p{p-q})\begin{pmatrix} \mathfrak{A} & \mathfrak{B} \\ \mathfrak{C} & \mathfrak{D} \end{pmatrix},
\end{equation}
where
\begin{subequations}
\label{FluMaComps}
\begin{align}
\mathfrak{A} =& -(\Psi^\dagger_\gamma\eta^X)^T(\Psi_\delta^\dagger\eta^Y)f(0,-p,0,p)\nonumber\\
&- (\Psi^\dagger_\gamma\eta^Y)^T(\Psi_\delta^\dagger\eta^X)f(0,p,0,-p),\\
\mathfrak{B} =& (\Psi^\dagger_\gamma\eta^X)^T(\eta^Y\Psi_\delta)^Tf(0,-p,-p,0)\nonumber\\
&+ (\Psi^\dagger_\gamma\eta^Y)^T(\eta^X\Psi_\delta)^Tf(-p,0,0,-p)\nonumber\\
&- {\eta^X}^T\delta_{\gamma\delta}(\Psi_\alpha^\dagger\eta^Y\Psi_\alpha)f(-p,-p,0,0)\nonumber\\
&- {\eta^B}^T\delta_{\gamma\delta}(\Psi_\alpha^\dagger\eta^X\Psi_\alpha)f(0,0,-p,-p),\label{BLine}\\
\mathfrak{C} =& (\eta^X\Psi_\gamma)(\Psi^\dagger_\delta\eta^Y)f(p,0,0,p)\nonumber\\
&+ (\eta^Y\Psi_\gamma)(\Psi^\dagger_\delta\eta^X)f(0,p,p,0)\nonumber\\
&+ \eta^A\delta_{\gamma\delta}(\Psi_\alpha^\dagger\eta^Y\Psi_\alpha)f(p,p,0,0)\nonumber\\
&+\eta^Y\delta_{\gamma\delta}(\Psi_\alpha^\dagger\eta^X\Psi_\alpha)f(0,0,p,p),\label{CLine}\\
\mathfrak{D} =& -(\eta^X\Psi_\gamma)(\eta^Y\Psi_\delta)^Tf(p,0,-p,0)\nonumber\\
&- (\eta^Y\Psi_\gamma)(\eta^X\Psi_\delta)^Tf(-p,0,p,0).
\end{align}
\end{subequations}
The flow equation~\eqref{Wettereq} may be expanded in powers of $\mathcal{F}_\Lambda$. The quadratic contribution 
\begin{equation}
\label{QuadTrace}
\sim\m{Tr}\left[\mathcal{P}_\Lambda^{-1}\mathcal{F}_\Lambda \mathcal{P}_\Lambda^{-1}\mathcal{F}_\Lambda\right]
\end{equation}
of this expansion determines the flow of the fermionic interaction terms themselves. 
Thus, any $\partial_\Lambda J_n \sim \frac{1}{N}$. Only terms where the trace operation on the right hand side of the Wetterich equation contributes an extra factor of $N$ avoid being suppressed in the limit $N\rightarrow\infty$.
By construction, $\mathcal{P}_\Lambda$ is diagonal in the spin indices. To achieve an overall $N = \m{Tr}[\delta_{\gamma\delta}]$, only components $\sim\delta_{\gamma\delta}$ in eqns.~\eqref{FluMaComps} can contribute. These are given in the second lines of eqns.~\eqref{BLine} and~\eqref{CLine}. Multiplying those terms according to eq.~\eqref{QuadTrace} can never result in any product of matrices $\sigma^x\cdot\eta^Y$ sandwiched between $\Psi^\dagger,\Psi$. Since this is the only way to generate new matrix structures besides the existing $\eta^{X/Y}$, it can be concluded that no such new structures can be generated at large $N$.

\section{Gell-Mann matrices of SU(4)}
\label{GellMaaan}
For definiteness, we here list the generators of SU(4) in the fundamental representation as used in Sec.~\ref{Sub2_FancyGroups}.
\begin{equation}
\hat{\lambda}_1 = 
\begin{pmatrix} 
0 & 1 & 0 & 0\\
1 & 0 & 0 & 0\\ 
0 & 0 & 0 & 0\\ 
0 & 0 & 0 & 0
\end{pmatrix},\quad
\hat{\lambda}_2 = 
\begin{pmatrix} 
0 & 0 & 1 & 0\\
0 & 0 & 0 & 0\\ 
1 & 0 & 0 & 0\\ 
0 & 0 & 0 & 0
\end{pmatrix},
\end{equation}

\begin{equation}
\hat{\lambda}_3 = 
\begin{pmatrix} 
0 & 0 & 0 & 0\\
0 & 0 & 1 & 0\\ 
0 & 1 & 0 & 0\\ 
0 & 0 & 0 & 0
\end{pmatrix},\quad
\hat{\lambda}_4 = 
\begin{pmatrix} 
0 & 0 & 0 & 1\\
0 & 0 & 0 & 0\\ 
0 & 0 & 0 & 0\\ 
1 & 0 & 0 & 0
\end{pmatrix},
\end{equation}

\begin{equation}
\hat{\lambda}_5 = 
\begin{pmatrix} 
0 & 0 & 0 & 0\\
0 & 0 & 0 & 1\\ 
0 & 0 & 0 & 0\\ 
0 & 1 & 0 & 0
\end{pmatrix},\quad
\hat{\lambda}_6 = 
\begin{pmatrix} 
0 & 0 & 0 & 0\\
0 & 0 & 0 & 0\\ 
0 & 0 & 0 & 1\\ 
0 & 0 & 1 & 0
\end{pmatrix},
\end{equation}

\begin{widetext}
\begin{equation}
\hat{\lambda}_7 = 
\begin{pmatrix} 
0 & -i & 0 & 0\\
i & 0 & 0 & 0\\ 
0 & 0 & 0 & 0\\ 
0 & 0 & 0 & 0
\end{pmatrix},\quad
\hat{\lambda}_8 = 
\begin{pmatrix} 
0 & 0 & -i & 0\\
0 & 0 & 0 & 0\\ 
i & 0 & 0 & 0\\ 
0 & 0 & 0 & 0
\end{pmatrix},\quad 
\hat{\lambda}_9 = 
\begin{pmatrix} 
0 & 0 & 0 & 0\\
0 & 0 & -i & 0\\ 
0 & i & 0 & 0\\ 
0 & 0 & 0 & 0
\end{pmatrix},\quad
\hat{\lambda}_{10} = 
\begin{pmatrix} 
0 & 0 & 0 & -i\\
0 & 0 & 0 & 0\\ 
0 & 0 & 0 & 0\\ 
i & 0 & 0 & 0
\end{pmatrix},
\end{equation}

\begin{equation}
\hat{\lambda}_{11} = 
\begin{pmatrix} 
0 & 0 & 0 & 0\\
0 & 0 & 0 & -i\\ 
0 & 0 & 0 & 0\\ 
0 & i & 0 & 0
\end{pmatrix},\quad
\hat{\lambda}_{12} = 
\begin{pmatrix} 
0 & 0 & 0 & 0\\
0 & 0 & 0 & 0\\ 
0 & 0 & 0 & -i\\ 
0 & 0 & i & 0
\end{pmatrix},\quad
\hat{\lambda}_{13} = 
\begin{pmatrix} 
1 & 0 & 0 & 0\\
0 & -1 & 0 & 0\\ 
0 & 0 & 0 & 0\\ 
0 & 0 & 0 & 0
\end{pmatrix},\quad
\hat{\lambda}_{14} = 
\frac{1}{\sqrt{3}}\begin{pmatrix} 
1 & 0 & 0 & 0\\
0 & 1 & 0 & 0\\ 
0 & 0 & -2 & 0\\ 
0 & 0 & 0 & 0
\end{pmatrix},
\end{equation}

\begin{equation}
\hat{\lambda}_{15} = 
\frac{1}{\sqrt{6}}\begin{pmatrix} 
1 & 0 & 0 & 0\\
0 & 1 & 0 & 0\\ 
0 & 0 & 1 & 0\\ 
0 & 0 & 0 & -3
\end{pmatrix}.
\end{equation}

\section{Flow equations of the toy model}
\label{ToyEquations}
Employing the regulator~\eqref{EnssReg} the flow equations for the toy model of Sec.~\ref{PopotovRulez} are given by
\begin{subequations}
\begin{align}
\partial_\Lambda\mu_\Lambda =& -\frac{2}{\pi}\frac{\omega_\Lambda^2\tilde{\vartheta}_\Lambda\mu_\Lambda}{(\omega_\Lambda^2 + \tilde{\vartheta}_\Lambda^2\mu_\Lambda^2)^2}(4L_\Lambda + X_\Lambda - 2J_\Lambda),\\
\partial_\Lambda J_\Lambda =& \frac{2}{\pi}\left[\frac{3\tilde{\vartheta}_\Lambda^3\mu_\Lambda^2\omega_\Lambda^2-\tilde{\vartheta}_\Lambda\omega_\Lambda^4}{(\tilde{\vartheta}_\Lambda^2\mu_\Lambda^2+\omega_\Lambda^2)^3}(J_\Lambda^2-J_\Lambda L_\Lambda - 2J_\Lambda X_\Lambda) - \frac{J_\Lambda L_\Lambda\tilde{\vartheta}_\Lambda\omega_\Lambda^2}{(\tilde{\vartheta}_\Lambda^2\mu_\Lambda^2+\omega_\Lambda^2)^2}\right]\nonumber\\
&+ 2T\sum_n(\partial_\Lambda\mu_\Lambda)\vartheta_\Lambda^4\mu_\Lambda\left[\frac{\vartheta_\Lambda^2\mu_\Lambda^2 - 3\omega_n^2}{(\omega_n^2+\vartheta_\Lambda^2\mu_\Lambda^2)^3}(J_\Lambda^2-J_\Lambda L_\Lambda-2J_\Lambda X_\Lambda) - \frac{J_\Lambda L_\Lambda}{(\omega_n^2+\vartheta_\Lambda^2\mu_\Lambda^2)^2}\right],\\
\partial_\Lambda L_\Lambda =& \frac{1}{\pi}\left[\frac{3\tilde{\vartheta}_\Lambda^3\mu_\Lambda^2\omega_\Lambda^2-\tilde{\vartheta}_\Lambda\omega_\Lambda^4}{(\tilde{\vartheta}_\Lambda^2\mu_\Lambda^2+\omega_\Lambda^2)^3}(4L_\Lambda X_\Lambda - L_\Lambda^2 - 4J_\Lambda X_\Lambda) - \frac{(J_\Lambda^2+L_\Lambda^2)\tilde{\vartheta}_\Lambda\omega_\Lambda^2}{(\tilde{\vartheta}_\Lambda^2\mu_\Lambda^2+\omega_\Lambda^2)^2}\right]\nonumber\\
&+ T\sum_n(\partial_\Lambda\mu_\Lambda)\vartheta_\Lambda^4\mu_\Lambda\left[\frac{\vartheta_\Lambda^2\mu_\Lambda^2 - 3\omega_n^2}{(\omega_n^2+\vartheta_\Lambda^2\mu_\Lambda^2)^3}(4L_\Lambda X_\Lambda - L_\Lambda^2 - 4J_\Lambda X_\Lambda) - \frac{J_\Lambda^2+L_\Lambda^2}{(\omega_n^2+\vartheta_\Lambda^2\mu_\Lambda^2)^2}\right],\\
\partial_\Lambda X_\Lambda =& \frac{2}{\pi}\left[\frac{3\tilde{\vartheta}_\Lambda^3\mu_\Lambda^2\omega_\Lambda^2-\tilde{\vartheta}_\Lambda\omega_\Lambda^4}{(\tilde{\vartheta}_\Lambda^2\mu_\Lambda^2+\omega_\Lambda^2)^3}(8L_\Lambda^2-4J_\Lambda^2-X_\Lambda^2-8J_\Lambda L_\Lambda) - \frac{X_\Lambda^2\tilde{\vartheta}_\Lambda\omega_\Lambda^2}{(\tilde{\vartheta}_\Lambda^2\mu_\Lambda^2+\omega_\Lambda^2)^2}\right]\nonumber\\
&+ 2T\sum_n(\partial_\Lambda\mu_\Lambda)\vartheta_\Lambda^4\mu_\Lambda\left[\frac{\vartheta_\Lambda^2\mu_\Lambda^2 - 3\omega_n^2}{(\omega_n^2+\vartheta_\Lambda^2\mu_\Lambda^2)^3}(8L_\Lambda^2-4J_\Lambda^2-X_\Lambda^2-8J_\Lambda L_\Lambda) - \frac{X_\Lambda^2}{(\omega_n^2+\vartheta_\Lambda^2\mu_\Lambda^2)^2}\right],
\end{align}
\end{subequations}
\end{widetext}
where 
\begin{equation}
\tilde{\vartheta}_{\Lambda} = 1+\left\lfloor \frac{\Lambda}{2\pi T}\right\rfloor - \frac{\Lambda}{2\pi T} 
\end{equation}
and
\begin{equation} 
\omega_\Lambda = \pi T\left(2\left\lfloor \frac{\Lambda}{2\pi T}\right\rfloor + 1\right).
\end{equation}
In the limit $T\rightarrow 0$, $\mu_\Lambda$ vanishes. The flow equations thus simplify dramatically, yielding
\begin{subequations}
\begin{align}
\partial_\Lambda J_\Lambda &= -\frac{J_\Lambda^2-2J_\Lambda X_\Lambda}{\pi \Lambda^2},\\
\partial_\Lambda L_\Lambda &= -\frac{4L_\Lambda X_\Lambda - 4J_\Lambda X_\Lambda + J_\Lambda^2}{2\pi \Lambda^2},\\
\partial_\Lambda X_\Lambda &= -\frac{8L_\Lambda^2 - 4J_\Lambda^2 - 8J_\Lambda L_\Lambda}{\pi \Lambda^2}\,,
\end{align}
\end{subequations}
We use these equations to compute the $T=0$ data point in Fig.~\ref{PopotovSSB}.
\newpage

%------------------------------------
% Bibliography
%------------------------------------

%------------------------------------
\bibliography{LargeMess}
%------------------------------------

%------------------------------------
\end{document}